\documentclass[11pt]{article}
\usepackage[toc,page]{appendix}
\usepackage[utf8x]{inputenc}
\usepackage[colorlinks]{hyperref}
\usepackage{amsmath}
\usepackage{amssymb}
\usepackage{amsthm}
\usepackage[frak=esstix]{mathalpha}
\usepackage{mathtools}
\usepackage{physics}
\usepackage{booktabs}
\usepackage{color,soul}
\usepackage{tocloft}
\usepackage{xcolor}
\usepackage{geometry} 
\usepackage{hyperref}
\usepackage{setspace}
\usepackage{fullpage}[cm] 
\usepackage[font=small,format=plain,labelfont=bf,up,textfont=normal,up]{caption}

\swapnumbers

\selectcolormodel{cmy}
\hypersetup{
	colorlinks,
	linkcolor = {cyan},
	citecolor = {green},
	urlcolor  = {magenta}
}

\newcommand\widebar[1]{\mathop{\overline{#1}}}

\newcommand{\diff}{\ensuremath{\textrm{diff}}}
\newcommand{\TT}{\ensuremath{{\textrm{T}\bar{\textrm{T}}}}}
\newcommand{\STT}{\ensuremath{S_{\textrm{T}\bar{\textrm{T}}}}}

\newcommand{\Pd}{\ensuremath{P^{\dagger}}}
\newcommand{\pd}{\ensuremath{\partial}}
\newcommand{\bpd}{\ensuremath{\bar{\partial}}}

\newcommand{\pr}[1]{\left(#1\right)}

\newcommand{\DD}{\ensuremath{\mathcal{D}}}
\newcommand{\OO}{\ensuremath{\mathcal{O}}}

\newcommand{\FF}{\ensuremath{\mathcal{F}}}
\newcommand{\ATS}{\ensuremath{\mathcal{A}}}
\newcommand{\ee}{\ensuremath{\varepsilon}}

\newcommand{\al}{\ensuremath{\alpha}}
\newcommand{\bb}{\ensuremath{\beta}}
\newcommand{\g}{\ensuremath{\gamma}}
\newcommand{\de}{\ensuremath{\delta}}
\newcommand{\Om}{\ensuremath{\Omega}}
\newcommand{\sg}{\ensuremath{\sigma}}
\newcommand{\ep}{\ensuremath{\epsilon}}
\newcommand{\La}{\ensuremath{\Lambda}}

\newcommand{\floor}[1]{\left\lfloor #1 \right\rfloor}

\newcommand{\sn}{\ensuremath{\displaystyle\sum_{i=1}^n}}

\newcommand{\pn}{\ensuremath{\displaystyle\prod_{i=1}^n}}

\begin{document}

\vspace{1.8cm}
\begin{center}        % Main title
  \Huge Correlation Functions in $\TT$-deformed Conformal Field Theories
\end{center}  

\vspace{0.7cm}
\begin{center}        % Authors
{\large  Ofer Aharony and Netanel Barel}
\end{center}

\vspace{0.15cm}
\begin{center}        % Institutes
\emph{Department of Particle Physics and Astrophysics, Weizmann Institute of Science, \\ Rehovot 7610001, Israel}\\[.4cm]
             
e-mail: \tt ofer.aharony@weizmann.ac.il, netanel.barel@weizmann.ac.il
\end{center}

\vspace{1.5cm}

\begin{abstract}

We study the correlation functions of local operators in unitary $\TT$-deformed field theories, using their formulation in terms of Jackiw-Teitelboim gravity. The position of the operators is defined using the dynamical coordinates of this formalism. We focus on the two-point correlation function in momentum space, when the undeformed theory is a conformal field theory. In particular, we compute the large momentum behavior of the correlation functions, which manifests the non-locality of the $\TT$-deformed theory. The correlation function has UV-divergences, which are regulated by a point-splitting regulator. Renormalizing the operators requires multiplicative factors depending on the momentum, unlike the behavior in local QFTs. The large momentum limit of the correlator, which is the main result of this paper, is proportional to $|q|^{-\frac{q^2}{\pi|\La|}}$, where $q$ is the momentum and $1/|\La|$ is the deformation parameter. Interestingly, the exponent here has a different sign from earlier results obtained by resummation of small $q$ computations. The decay at large momentum implies that the operators behave non-locally at the scale set by the deformation parameter.
 
\end{abstract}

\thispagestyle{empty}
\clearpage

\tableofcontents

\setcounter{page}{1}

% ====================================
\section{Introduction}\label{sec_int}

In recent years it was discovered that in $1+1$ dimensional space-time, special types of irrelevant deformations of quantum field theories (QFTs) are under control and even solvable \cite{Smirnov:2016lqw,Cavaglia:2016oda}. One of them is the $\TT$-deformation, which is the theory investigated in this paper (a review of this topic with a complete list of references can be found in \cite{Jiang:2021kfo}).
The $\TT$-operator is defined as the determinant of the energy-momentum tensor, using the following normalization:
\begin{equation}\label{def_EMT}
    T^{\al\bb} = \frac{2}{\sqrt{g}}\frac{\de S}{\de g_{\al\bb}}\Big|_{g = \de},
\end{equation}
where Euclidean signature is used, $\sg^{1,2}$ are coordinates in this space, and $\de$ is the flat metric. We will often use holomorphic and anti-holomorphic coordinates:
\begin{equation}\label{compcoor}
    w \equiv \sg^1 + i\sg^2 \ ,\qquad \bar{w} \equiv \sg^1 - i\sg^2.
\end{equation}
Using these coordinates, the following quantities are defined:
\begin{equation}
    T \equiv T_{ww},\qquad \bar{T} \equiv T_{\bar{w}\bar{w}},\qquad \Theta \equiv T_{w\bar{w}}.
\end{equation}
Then, the $\TT$-operator is given by:
\begin{equation}\label{def_TT_1}
    \TT \equiv \det\left( T_{\al\bb} \right) = -4\left( T_{ww}T_{\bar w \bar w} - T^2_{w\bar w} \right) = -4\left(T \bar T - \Theta^2 \right).
\end{equation}
$\TT$ denotes this composite operator, which is different from the product $T\bar{T}$ in general. One can show that, when this operator is integrated over space, there is no short-distance singularity when bringing together the different $T$'s and $\Theta$'s in \eqref{def_TT_1}, so their product at the same point is well-defined. The $\TT$-operator has dimension of mass$^4$, so its deformation comes with a coefficient $t$ with dimension of mass$^{-2}$=length$^2$.
For each value of $t$, one defines the Lagrangian of the deformed theory at $t+dt$ in terms of the deformed theory at $t$:
\begin{equation}\label{def_TT_2}
    \mathcal{L}^{(t+dt)} = \mathcal{L}^{(t)} + \TT^{(t)}dt,
\end{equation}
where $\TT^{(t)}$ is obtained using the energy-momentum tensor derived from $S^{(t)}$. It is easy to see that Lorentz-invariance is maintained
if $\mathcal{L}^{(0)}$ is Lorentz invariant. The deformation was found to be unitary for $t\geq0$, i.e., the energies are real\footnote{at least at large enough volume.}. This is opposed to the case when $t<0$, where some energies are complex, and therefore, only the positive $t$ case will be considered here. We note that the coefficient in the definition of the $\TT$-operator in \eqref{def_TT_1} and the coefficient in the definition of the flow in \eqref{def_TT_2} are different from paper to paper, and one has to be careful when comparing results.

Not much is known about the $\TT$-deformation. The energy levels of the theory on a circle, and thus the torus partition function, are explicitly known for any $t$, as well as the S-matrix. The definition of the deformation presented above is in a sense perturbative, and by perturbative methods the two-point correlation function of a deformed CFT was calculated. There were calculations to leading order in the small deformation parameter \cite{Kraus:2018xrn,Giribet:2017imm,Aharony:2018vux,He:2020qcs,He:2019vzf,He:2019ahx,Ebert:2020tuy,He:2022jyt,He:2020udl}, with specific attention to the correlation functions of the components of the energy-momentum tensor \cite{Rosenhaus:2019utc,Li:2020pwa,Hirano:2020ppu,He:2023hoj}. The two-point correlation function of general operators for a general deformation parameter was calculated at large momentum in \cite{Cardy:2019qao}. It was found to be:
\begin{equation}\label{cor_Crdy}
    C^t(q) = C^0(q)\left(\frac{|q|}{\mu}\right)^{\frac{tq^2}{\pi}},
\end{equation}
where $2\pi\lambda$ in \cite{Cardy:2019qao} is $t$ here. This result was also derived using other methods in \cite{Hirano:2020nwq}. In this equation, $C^0(q)$ is the undeformed CFT correlation function, which is some power law $C^0(q)=|q|^{2\Delta-2}$, $\Delta$ being the dimension of the operators in the undeformed theory, and $\mu$ is a renormalization scale. This result was obtained using a momentum-dependent renormalization of the operators, unlike in local QFTs. 

It was suggested in \cite{Dubovsky:2017cnj} that the $\TT$-deformed theories can be described non-perturbatively using the flat space Jackiw-Teitelboim (JT) gravity \cite{Dubovsky:2017cnj}. Using the JT-description the S-matrix, the spectrum on a circle and the partition function on the torus were reproduced \cite{Dubovsky:2017cnj,Dubovsky:2018bmo}. 
In this paper, the correlation functions of the deformed theory will be calculated, using the JT-formulation, which is non-perturbative in the deformation parameter. It will be shown that in this formulation the need for a similar momentum-dependent renormalization also shows up. However, we will see that the precise dependence of the two-point correlator on the momentum at large momentum is very different from \eqref{cor_Crdy}, and it is proportional to
\begin{equation}\label{hml_6n}
    \frac{1}{\sin\left(\frac{t q^2}{2} + \pi\Delta\right)} \left(\frac{1}{\mu t|q|}\right)^{\frac{t q^2}{\pi}+2\Delta}.
\end{equation}
The interpretation of this result, and its relation to \eqref{cor_Crdy}, will be discussed in the final section. Very recently, the same correlation function \eqref{hml_6n} was found for a string theory correlation function in \cite{Cui:2023jrb}, and the relation to that result will also be discussed in the final section.

In the JT-description, the action is given by:
\begin{equation}\label{INT_4}
    \STT(\psi,g_{\al\bb},\varphi) = S_0(\psi,g_{\al\bb}) + \int d^2\sg\sqrt{g}\left(\varphi R -\Lambda\right),
\end{equation}
where $\psi$ represents the field content of the undeformed theory, and $S_0(\psi,g_{\al\bb})$ is the action of the original field theory with a coupling to a dynamical metric $g_{\al\bb}$, and with no direct coupling to the dilaton field $\varphi$. For a general QFT, the coupling to the metric may be given by the minimal coupling. However, in the case where the undeformed theory is a CFT, as in this paper, the coupling will be chosen such that $S_0(\psi,g_{\al\bb})$ has Weyl invariance (the path-integral, however, may have a Weyl anomaly). The  energy density $\La$ in the vacuum (having dimensions mass$^2$), is connected to the $\TT$-deformation parameter $t$ by the relation:
\begin{equation}\label{INT_5}
    t = -\frac{1}{\Lambda}.
\end{equation}
In this formalism, the unitary theory is found for $\La=-|\La|<0$.  The undeformed theory is recovered in the limit $|\La|\to\infty$.

The JT-gravity action \eqref{INT_4} has an equivalent formulation using vielbein variables $e^a_{\al}$ and a spin-connection $\omega_{\al}$, instead of the metric components $g_{\al\bb} = e^a_\al e^b_\bb \de_{ab}$, where one introduces Lagrange multipliers $\lambda^a$ to force the spin-connection $\omega_{\al}$ to be compatible with the metric.  Defining $X^a\equiv\Lambda^{-1}\ep^a_b\lambda^b$, simple manipulations show that the action takes the form:
\begin{equation}\label{STT_action}
\begin{split}
    \STT(\psi,e^a_{\al},X^a) & = S_0(\psi,e^a_{\al}) + S_{JT}(e^a_{\al},X^a) = \\
    & = S_0(\psi,e^a_{\al}) +
    \frac{\Lambda}{2}\int d^2\sg \ep^{\al\bb}\ep_{ab}(\partial_{\al}X^a-e^a_{\al})(\partial_{\bb}X^b-e^b_{\bb}).
    \end{split}
\end{equation} 
The advantage of this description relative to the former, is that the $X^a$ play the role of dynamical coordinates of the target-space (TS), which is identified with the original space of the undeformed theory \cite{Dubovsky:2017cnj,Dubovsky:2018bmo}. In contrast, the worldsheet (WS) in which the action \eqref{STT_action} is formulated, parameterized by coordinates $\sg$, is viewed as an auxiliary space. Therefore, one can introduce local operators on the TS, and treat them as a deformation of the undeformed operators.\footnote{It is less natural to consider local operators on the WS, since the theory includes a dynamical metric on the WS and the notion of a local operator is subtle.} Hence, the following definition for local operators will be adopted:
\begin{equation}\label{op_def_pos}
    \OO(X_0) = \int d^2\sg\sqrt{g(\sg)} O(\sg)\de\left(X(\sg)-X_0\right),
\end{equation} 
where $O(\sg)$ is some operator in the undeformed theory. Alternatively, one can replace $\OO(X_0)$ by its Fourier transform to avoid the $\de$-function of an operator. In Fourier space we have: 
\begin{equation}\label{op_def_mom}
    \OO(q_0) = \int d^2\sg\sqrt{g(\sg)} O(\sg)\exp\left(iq_0\cdot X(\sg)\right).
\end{equation}
The correlation function in this formalism is defined as follows:
\begin{equation}\label{def_cor}
\begin{split}
    C(q_1,q_2) \equiv& \left<\OO(q_1)\OO(q_2)\right> = \frac{1}{Z_{\TT}}\int\frac{\DD e\DD X\DD\psi}{V_\diff}\OO(q_1)\OO(q_2)e^{-\STT} \\
    = & \frac{1}{Z_{\TT}}\int\frac{\DD e\DD X\DD\psi}{V_\diff}\int d^2\sg_1d^2\sg_2 \sqrt{g(\sg_1)}O(\sg_1)\sqrt{g(\sg_2)}O(\sg_2)e^{iq_1\cdot X(\sg_1)+iq_2\cdot X(\sg_2)-\STT},
\end{split}
\end{equation}
where the partition function is:
\begin{equation}
        Z_{\TT} \equiv \int\frac{\DD e\DD X\DD\psi}{V_\diff}e^{-\STT}.
\end{equation}
We will calculate the two-point correlator \eqref{def_cor} on the plane, and study its behavior as a function of the momentum.\footnote{$Z_{\TT}$ was already calculated in \cite{Dubovsky:2018bmo}. It will be omitted, as is does not depend on the momentum.} In particular, we will be interested in the large momentum limit of this correlator, and use it to learn about locality properties of the $\TT$-deformed field theories. In the usual case, where the UV limit of a QFT is a CFT, in the large momentum limit the correlation function behaves as a power law $|q|^{2\Delta-2}$, and it represents the short distance limit. As a result, it will the same for any two-dimensional manifold on which we put the QFT (for example, the torus). We will find in the $\TT$-deformed theory a different behavior than that of a CFT, which indicates its non-local nature. Given the non-locality, another natural question is to compare the calculation on an infinite plane to that on a torus. This question will not be answered here, although the calculation on a torus is similar in many ways to the one that  will be presented here.  

The structure of this paper is the following. In section \ref{sec_pro} the general properties of the JT-gravity theory will be described. In section \ref{sec_PI} we gauge-fix the diffeomorphism (diff) invariance by choosing a specific coordinate system. We explain in detail the meaning of the gauge chosen and the procedure of calculating the Faddeev-Popov (FP) determinant in appendix \ref{sec_FP}. Some global aspects of the coordinate system chosen are discussed in appendix \ref{sec_boun}. Then, some formal manipulations will be done on the gauge-fixed expression. In section \ref{sec_reg} regularization and renormalization procedures will be carried out to make the expression well-defined and finite. Then, the main result, which is the large momentum limit of the two-point correlator, will be derived. Appendix \ref{sec_fur} completes the calculation made in this section. In the last section, section \ref{sec_dis}, we compare the large momentum result to previous results and conclude. In appendix \ref{sec_hgh_lam} the small momentum limit of the correlator is calculated, and the result is compared to the perturbative calculation in the $\TT$-formulation. Finally, in appendix \ref{sec_diff_op} an alternative definition for operators in the JT-formalism is considered, and it is shown that they share the same large momentum behavior as for the operators discussed along the main text.

% ====================================
\section{General Properties of the Action}\label{sec_pro}

In this section some properties of the JT-action \eqref{STT_action} are discussed. TS indices will be written using Latin letters and WS indices using Greek letters. With respect to the WS and the TS, the field content in the action has the following properties:
\begin{enumerate}
    \item $\psi$ - The fields of the undeformed theory. They have some structure with respect to the WS, and we consider some primary operator $O$ which is built using them. They are scalars with respect to the TS. 
    
    \item $X^a$ - Coordinates on the TS. As such, they are scalars of the WS. Because the TS is flat, the coordinates behave as a vector. We are interested in mappings from the WS to the TS which cover it once, so it will be convenient to write $X^a(\sg)$ in the form:
    \begin{equation}\label{rel_X_Y_p}
        X^a(\sg)\equiv I^a_{\al}\sg^{\al}+Y^a(\sg)
    \end{equation}
    (where $I^a_\al$ is a constant matrix whose values are $I^1_1=I^2_2=1,\ I^1_2=I^2_1=0$).
    The first term in \eqref{rel_X_Y_p} describes passing over the whole TS, the analogue of ``winding one'' on a torus, and will be called the winding part. The second term describes local changes from it, which are finite as $|\sg|\to\infty$.
    
    \item $e^a_{\al}$ - The vielbein. They are vectors of the TS and co-vectors (1-form) of the WS. 
\end{enumerate}

We add here three comments. The first concerns the limit $|\La|\to\infty$, where the undeformed correlation function should be recovered. In this limit, one can use a saddle point approximation for $S_{JT}$ in \eqref{STT_action}, and the path-integral (PI) over $\DD e\DD X$ is trivial (after an appropriate gauge-fixing). Therefore, one has to replace any appearance of $e^a_\al,X^a$ by their values at the saddle point (``saddle"): 
\begin{equation}
\begin{split}
    e^a_\al(\sg) &\rightarrow I^a_\al, \\
    X^a(\sg) &\rightarrow I^a_\al\sg^\al.   
\end{split}
\end{equation}
Then, one gets the following limits:
\begin{equation}
\begin{split}
    Z_{\TT} &\rightarrow \int\DD\psi e^{-S_0(\psi,\de)+0}=Z_0, \\
    \OO(q) &\rightarrow \int d^2\sg\cdot 1\cdot O(\sg)e^{iq\cdot \sg}=O(q),
\end{split}
\end{equation}
and the two-point correlator reduces to the momentum space correlation function of the undeformed theory, as expected.

The second comment is about regularization and renormalization. The theory will not be regularized from the beginning. Instead, progress in formal developments will be made, as far as possible, and only after gathering all the divergent pieces, it will be regularized. In this way, the calculations are much easier, and it will be clear what is the appropriate UV-regularization to use. It will be chosen such that the assumptions made while deriving the formal expression will be justified. As for IR-regularization, one way to do it is to put the theory on a torus. However, in this paper, for simplicity, we will not go into any details of this regularization, but just use its general features.

The third comment concerns the decomposition \eqref{rel_X_Y_p}. By using this decomposition, we assume specific boundary conditions for the $X^a(\sg)$ and for allowed diffeomorphisms $\sg'(\sg)$ at $|\sg|\to\infty$. In particular, the allowed diffeomorphisms should go to the identity map plus some bounded function, and $Y^a(\sg)$ should have a Fourier transform. We could also use boundary conditions for the vielbein $e^a_\al(\sg) \to I^a_\al$ as $|\sg|\to\infty$. However, we will be more general, and require only that the WS should be flat at the boundary, i.e., constant vielbein. In the infinite plane limit, it will become $I^a_\al$ dynamically. The boundary conditions for $\psi(\sg)$ are those used in the undeformed CFT. We refer the reader to appendix \ref{sec_boun} for more details.

% ====================================
\section{Path Integral Calculation}\label{sec_PI}

The diffeomorphism freedom in the computation \eqref{def_cor} will be gauge-fixed following the usual FP procedure. We choose the orthogonal gauge (sometimes called the conformal gauge), where the vielbein takes the form:  
\begin{equation}\label{fx_vln_sg_pl}
    \bar{e}^a_{\al}({\sg}) \equiv e^{\Om(\sg)}(e^{\ep \phi(\sg)})^\bb_\al I^a_\bb \ \footnote{Since $e^{\Om(\sg)}$ is just a WS scalar, and $e^{\ep\phi(\sg)}$ lives in the local frame, and is therefore a WS scalar as well, only the explicit matrix $I^a_\al$ carries a WS index and transforms under a change of coordinates. In particular, one set of two orthogonal vectors is $I_\al^a$ (all other sets are rotations of these two by an angle $\phi(\sg)$ at each point $\sg$).}
    \qquad \Rightarrow \qquad \bar{g}_{\al\bb}({\sg}) = e^{2\Om(\sg)}\de_{\al\bb},
\end{equation}
and we denote:
\begin{equation}\label{def_ep}
    \ep^\bb_\al \equiv
    \begin{pmatrix} 0 & 1 \\ -1 & 0 \end{pmatrix} \rightarrow
    \left( e^{\ep\phi} \right)^\bb_\al =
    \begin{pmatrix} \cos(\phi) & \sin(\phi) \\ -\sin(\phi) & \cos(\phi) 
    \end{pmatrix}.
\end{equation}
It should be noted that the generally covariant action \eqref{STT_action} does not have Weyl invariance (i.e., $e^a_{\al}(\sg)\rightarrow e^{\Om(\sg)}e^a_{\al}(\sg)$ cannot be compensated by other manipulations). Thus, after removing the diff-invariance redundancy by gauge-fixing the metric, the metric will remain with one degree of freedom ($3-2=1$). In the orthogonal gauge used here, this is the scaling factor $e^{2\Om(\sg)}$. As for the vielbein, except for the scaling factor $e^{\Om(\sg)}$, which the vielbein inherits from the metric, one has also their local orientation $\phi(\sg)$. As usual, the orthogonal gauge choice leaves the freedom of conformal transformations, but most of these are not allowed by the boundary conditions; only the freedom of translations remains. They will be fixed by taking the coordinate of the first operator insertion to $\sigma_1^\al = 0$.

Then, we use the FP identity for the integral over the space of diffeomorphisms $V$:
\begin{equation}\label{FP_id_3}
    1 = \int\DD V\DD\Om\DD\phi J(e,\sg_1) \de\left(e^{(V)} - \bar e(\Om,\phi) \right)\de\left(\sg_1^{(V)}-0\right).
\end{equation}
The determinant $J(\bar e,\sg_1)$ can be calculated, and it is $\sqrt{\det(c)}$, with:
\begin{equation}
    c^{\de}_{\bb} = -\frac{1}{2}{\de}^{\de}_{\bb}\pd^2 + \pd^{\de}\Om\pd_{\bb} - \pd_{\bb}\Om\pd^{\de} - {\de}^{\de}_{\bb}\pd_{\al}\Om\pd^{\al} + \frac{\de(\sg-\sg_1)}{\sqrt{\bar{g}(\sg_1)}}\de^\de_\bb.
\end{equation} 
Using the complex coordinates, the determinant $J(\bar e,0)$ takes the simple form:
\begin{equation}\label{det_dec_1}
    J(\bar e,0) = |\det\left( e^{-2\Om}\right)| \left|\det(-\pd\bpd-4\pd\Om\bpd + \de(w))\right|, 
\end{equation}
where:
\begin{equation}
\begin{split}
    & \pd \equiv \frac{\pd}{\pd w}, \ \bpd \equiv \frac{\pd}{\pd \bar w},
    \\
    & d^2w \equiv 2 d\sg^1 d\sg^2, \ \de(w) \equiv \frac{\de\pr{\sg^1,\sg^2}}{2} \Rightarrow \int d^2w\de(w) = 1.
\end{split}
\end{equation}
The details of the calculation are left for appendix \ref{sec_FP}. Inserting \eqref{FP_id_3} into \eqref{def_cor} gives the following expression:
\begin{equation}\label{cor_dev_2}
    \begin{split}
        \int &\frac{\DD V\DD e\DD X \DD\psi\DD\Om\DD\phi dA(\sg_1)dA(\sg_2)}{V_\diff} \de\left(e^{(V)} -\bar e(\Om,\phi)\right)\de\left(\sg_1^{(V)}\right) \cdot \\  &\qquad\qquad J(e,\sg_1)O(\psi(\sg_1))O(\psi(\sg_2))e^{iq_1\cdot X(\sg_1)+iq_2\cdot X(\sg_2)-\STT(\psi,e,X)},
    \end{split}
\end{equation}
where the WS area element $dA(\sg) \equiv d^2\sg \sqrt{g(\sg)}$ is used.\footnote{In \eqref{cor_dev_2}, one integrates over the orientation-preserving diffeomorphisms and vielbeins. If the undeformed theory has a reflection symmetry then taking all diffeomorphisms and all vielbeins will give an overall factor of two. If it is not symmetric under reflection, it can be coupled only to orientation-preserving vielbeins.}
As usual, we allow ourselves to interchange the order of integration variables everywhere\footnote{assuming the theory is well-defined by some UV- and IR-regularizations, as mentioned.}.

% ------------------------------------
\subsection{The Vielbein and Diffeomorphism Path Integral}\label{subsec_PI_EV}

Next, we follow the usual FP procedure. The action $\STT$ and the operators $\OO(q)$ are diff-invariant, as is the FP determinant $J(\bar e,0)$. 
Using this fact, we can rewrite \eqref{cor_dev_2} as:
\begin{equation}\label{cor_dev_3}
    \begin{split}
        \int&\frac{\DD V\DD e\DD X \DD\psi\DD\Om\DD\phi dA(\sg_1)dA(\sg_2)}{V_\diff}\de\left(e^{(V)} -\bar e(\Om,\phi)\right)\de\left(\sg_1^{(V)}\right) \cdot \\
        & J\pr{e^{(V)},\sg_1^{(V)}} O\pr{\psi^{(V)}(\sg_1^{(V)})} O\pr{\psi^{(V)}(\sg_2^{(V)})} e^{iq_1\cdot X^{(V)}(\sg_1^{(V)}) + iq_2\cdot X^{(V)}(\sg_2^{(V)}) -\STT\pr{\psi^{(V)},e^{(V)},X^{(V)}}}.
    \end{split}
\end{equation}
We can now change variables (without changing the measure) to $\tilde{e} = e^{(V)}$, $\tilde X = X^{(V)}$, and similarly for $\psi$, $\sigma_1$ and $\sigma_2$, to obtain
\begin{equation}\label{cor_dev_4}
\begin{split}
    \int\frac{\DD V\DD \Tilde{e}\DD\Tilde{X} \DD\Tilde{\psi}\DD\Om\DD\phi dA(\tilde{\sg}_1) dA(\tilde{\sg}_2)}{V_\diff} & \de\left(\Tilde{e}-\bar{e}(\Om,\phi)\right)\de(\Tilde{\sg}_1)
    J(\Tilde{e},\Tilde{\sg}_1) \\ & O\pr{\tilde{\psi}(\tilde{\sg}_1)} O\pr{\tilde{\psi}(\tilde{\sg}_2)} e^{iq_1 \cdot {\tilde X}(\tilde{\sg}_1) + iq_2 \cdot {\tilde X}(\tilde{\sg}_2)} e^{-\STT(\Tilde{\psi},\Tilde{e},\Tilde{X})}.
    \end{split}
\end{equation}
Since now nothing depends on $V$ in the integrand, we can integrate $\DD V$ to give $V_{\diff}$, cancelling the $V_{\diff}$ in the denominator. Because of the $\de$-function, the $\Tilde{e}$ integration is also trivial, and we replace $\Tilde{e}$ by $\bar{e}$ everywhere. Therefore, \eqref{cor_dev_4} reduces to the following PI:
\begin{equation}\label{cor_dev_5}
    \begin{split}
        \int &\DD X \DD\psi\DD\Om\DD\phi dA(\sg_1)dA(\sg_2)\de(\sg_1) J(\bar{e}(\Om,\phi),0) \cdot \\
        & \qquad\qquad O(\psi(\sg_1))O(\psi(\sg_2))e^{iq_1\cdot X(\sg_1)+iq_2\cdot X(\sg_2)-S_0(\psi,\bar{e}(\Om,\phi))-S_{JT}(\bar{e}(\Om,\phi),X)},
    \end{split}
\end{equation}
where we relabeled the integration variables by removing the tildes. The trivial integral over $d^2\sg_1$ is not evaluated yet, to keep the symmetry between $\sg_1$ and $\sg_2$. 

% ------------------------------------
\subsection{The Target-Space Coordinates Path Integral}\label{subsec_PI_TS}

The next PI to be evaluated is the $X$ PI, because of its simple appearance in the integrand. Before doing so, we decompose $S_{JT}$ in \eqref{STT_action} into its parts, to see which are relevant for the $X$ PI. Using the decomposition \eqref{rel_X_Y_p}, $S_{JT}/\Lambda$ separates into six terms:
\begin{enumerate}
    
    \item $\frac{1}{2}\int d^2\sg \ep^{\al\bb} \ep_{ab} \bar e^a_{\al} \bar e^b_{\bb} = \int d^2\sg \bar e = \int d^2\sg\sqrt{\bar g}\equiv \bar A$, where $\bar e \equiv \det(\bar e^a_\al)$. This is the WS area. 
    
    \item $\frac{1}{2}\int d^2\sg \ep^{\al\bb}\ep_{ab}I^a_{\al}I^b_{\bb} = \int d^2\sg \equiv \ATS$. This is the TS area (the metric on the TS is taken to be the identity matrix). This term and the previous term are IR-divergent. 
    
    \item $\frac{1}{2}\int d^2\sg \ep^{\al\bb}\ep_{ab}\pd_{\al}Y^a\pd_{\bb}Y^b=\frac{1}{2}\int d^2\sg \ep^{\al\bb}\ep_{ab}\pd_{\al}(Y^a\pd_{\bb}Y^b)$ is a boundary term. On the plane it is not well-defined, since $Y^a$ does not necessarily vanish at infinity. However, on the torus, the $Y^a$ are periodic functions, and this term vanishes. Thus, we ignore this term. The action is then linear in $Y^a$, and they play the role of Lagrange multipliers.
    
    \item $\int d^2\sg \ep^{\al\bb}\ep_{ab}\pd_{\al}Y^aI^b_{\bb}$ - this is also a boundary term and will be treated as the former.
    
    \item $-\int d^2\sg \ep^{\al\bb}\ep_{ab} \bar e^a_{\al}I^b_{\bb}$.
    
    \item $-\int d^2\sg \ep^{\al\bb}\ep_{ab} \bar e^a_{\al}\pd_{\bb}Y^b$ - this term alone will contribute to the EOM of $Y^a$ (or $X^a$), which will be discussed below. 
    
\end{enumerate}
We change the $X^a(\sg)$ variables to $Y^a(\sg)$. The measure satisfies $\DD X=\DD Y$. The momentum-dependent and/or $Y$-dependent terms are: 
\begin{equation}\label{X_PI_2}
    e^{iq_1\cdot \sg_1+iq_2\cdot \sg_2} \int\DD Y e^{iq_1\cdot Y(\sg_1)+iq_2\cdot Y(\sg_2) + \La\int d^2\sg \ep^{\al\bb}\ep_{ab}\bar{e}^a_{\al}(\sg)\pd_{\bb}Y^b(\sg)}.
\end{equation}
We also need to keep the terms from $e^{-\STT}$ that are independent of $Y$,
\begin{equation}\label{Ie_term}
    e^{|\La|\bar A} e^{|\La|\ATS}
    e^{\La\int d^2\sg \ep^{\al\bb}\ep_{ab}I^a_{\al}\bar{e}^b_{\bb}}.
\end{equation}

The integral over $Y$ in \eqref{X_PI_2} is strictly speaking divergent, because each $Y^b(\sg)$ in the exponent multiplies the real quantity $-\ep^{\al\bb}\ep_{ab}\pd_{\bb}\bar{e}^a_{\al}(\sg)$. This means that the theory is not well-defined in the form presented. A simple analytic continuation of $Y$ will not solve the problem, because of the imaginary terms $i(q_1\cdot Y(\sg_1)+q_2\cdot Y(\sg_2))$. To overcome this difficulty, one should make the whole exponent imaginary. This can be done by Wick rotating $\bar{e}$, i.e $e^{\Om}\rightarrow ie^{\Om}$. In terms of $\Om,\phi$, it sends $\Om\rightarrow\Om+i\pi/2$, and thus, it does not change the measure $\DD\Om\DD\phi$.

Evaluation of the measure $\DD Y$ requires an inner product on the tangent space. The inner product $G_s$ presented for the WS scalars $\DD\Om,\DD\phi$ in appendix \ref{sec_FP} in \eqref{in_prod_sve} is suitable for the $Y^a$ also, treating them as two (WS) scalars:
\begin{equation}\label{in_prod_y}
    G_Y(\de Y^a,\de Z^b) \equiv |\La|\int d\bar A \de_{ab}\de Y^a \de Z^b = |\La|\int d^2\sg\sqrt{\bar g} \de_{ab}\de Y^a \de Z^b,
\end{equation}
where $d\bar A \equiv \sqrt{\bar{g}}d^2\sg$ is the WS area element induced by the orthogonal gauged metric $\bar{g}_{\al\bb}$. Then, formally $\DD Y = \displaystyle \prod_{a=1}^{2} \prod_{\sg=1}^N \sqrt{|\La|} \sqrt[4]{\bar g(\sg)} dY^a(\sg) = \det\left(|\La|e^{2\Om}\right)  \prod_{a=1}^{2} \prod_{\sg=1}^N dY^a(\sg)$, when we treat, for the formal development, the continuously infinite variables $Y^a(\sg)$ as $2N$ discrete variables. The integral over $Y$ then gives a multiplication of $\de$-functions:
\begin{equation}
\begin{split}
    & \prod_{b=1}^2 \prod_{\sg=1}^{N} (2\pi)\de \left( (q_1)_b\de_{\sg,\sg_1} + (q_2)_b\de_{\sg,\sg_2} - \La \ep^{\al\bb}\ep_{ab}\pd_{\bb}\bar{e}^a_{\al}(\sg) \right) = \\
    & \left(\det(2\pi)\right)^2 \de(q_1+q_2) \prod_{b=1}^2 \prod_{\sg=1}^{N-1} \de \left( (q_1)_b\de_{\sg,\sg_1} + (q_2)_b\de_{\sg,\sg_2} - \La \ep^{\al\bb}\ep_{ab}\pd_{\bb}\bar{e}^a_{\al}(\sg) \right).
\end{split}
\end{equation}
In the first line the last term was integrated by parts. In the second line, the derivative was considered as a difference. Then, summation over all $\sg$ yielded $\de\left((q_1)_b + (q_2)_b\right)$, which is momentum conservation, expected from TS translation symmetry. The momentum conservation replaced two $\de$-functions from the infinite multiplication, reducing $N$ to $N-1$ (which $\sg$'s were removed is not important for the following). The $\det(|\La|)\left(\det(2\pi)\right)^2$ will be ignored since it is just an infinite number\footnote{and also will be cancelled by a similar factor from the partition function.}.

The remaining terms of $S_{JT}$ after the analytic continuation can be written explicitly. The first term is:
\begin{equation}\label{ee_term}
    e^{-\frac{\La}{2}\int d^2\sg \ep^{\al\bb}\ep_{ab}\bar{e}^a_{\al}\bar{e}^b_{\bb}} \xrightarrow[]{\text{a.c.}} e^{\frac{\La}{2}\int d^2\sg \ep^{\al\bb}\ep_{ab}\bar{e}^a_{\al}\bar{e}^b_{\bb}} =  e^{-|\La|\bar{A}},
\end{equation}
where a.c. stands for analytic continuation (of $\bar{e}$). The second term, which appears in \eqref{Ie_term}, acquires an additional $i$. Putting all together, \eqref{cor_dev_5} becomes:
\begin{equation}\label{cor_dev_6}
        \begin{split}
        e^{|\La|\ATS}\de(q_1+q_2)\int &\DD\psi\DD\Om\DD\phi d\bar A(\sg_1)d\bar A(\sg_2)\de(\sg_1) J(\bar{e},0) \det\left(e^{2\Om}\right)\prod_{b=1}^2 \prod_{\sg=1}^{N-1}\de(K_b(\sg)) \cdot \\
        &  O(\sg_1)O(\sg_2) e^{iq_1\cdot \sg_1 +iq_2\cdot \sg_2-S_0(\psi,\bar{e}) - |\La|\left(\bar{A}+i\int d^2\sg\bar{e}^a_\al I^\al_a\right)}.
    \end{split}
\end{equation}
where $K_b(\sg) \equiv (q_1)_b\de_{\sg,\sg_1} + (q_2)_b\de_{\sg,\sg_2} - \La \ep^{\al\bb}\ep_{ab}\pd_{\bb}\bar{e}^a_{\al}(\sg)$.

% ------------------------------------
\subsection{Conformal Factor and Orientation Path Integral}\label{subsec_PI_OP}

The PI over $\Om$ and $\phi$ is simple because of the $\de(K_b)$, and will be evaluated next. As for the $Y$ PI, a measure on the $\Om,\phi$ function space is needed for evaluation of $\DD\Om\DD\phi$. The inner product defined in appendix \ref{sec_FP} in \eqref{in_prod_UE} will be used. Using this inner product, $\DD\Om = \prod_{\sg=1}^N \sqrt{|\La|}\sqrt[4]{\bar g(\sg)}d\Om(\sg)$ and the same for $\DD\phi$, hence $\DD\Om\DD\phi = \det\left(e^{2\Om}\right)  \prod_{\sg=1}^N d\Om(\sg)d\phi(\sg)$ (where again a factor of $\det(|\La|)$ was ignored). $\de(K_b)$ gives a differential equation for $\bar e^a_\al(\Om,\phi)$, i.e., for $\Om,\phi$: 
\begin{equation}\label{OP_PI_1}
    \ep^{\al\bb} \ep_{ab}\pd_{\al} \bar e^b_{\bb}(\sg) = -\frac{(q_1)_a}{|\La|}\de(\sg-\sg_1) -\frac{(q_2)_a}{|\La|}\de(\sg-\sg_2).
\end{equation}
In components, \eqref{OP_PI_1} reads:
\begin{equation}\label{OP_PI_2}
    \begin{pmatrix}
        \pd_{1}\bar{e}^2_2-\pd_{2}\bar{e}^2_1 \\ 
        -\pd_{1}\bar{e}^1_2+\pd_{2}\bar{e}^1_1
    \end{pmatrix}
    =
    -\frac{1}{|\La|}
    \begin{pmatrix}
        (q_1)_1 \\ (q_1)_2
    \end{pmatrix}
    \de(\sg-\sg_1)
    -\frac{1}{|\La|}
    \begin{pmatrix}
        (q_2)_1 \\ (q_2)_2
    \end{pmatrix}
    \de(\sg-\sg_2).
\end{equation}
Hence, on the flat coordinate space one has two copies of Gauss' law, one for each TS coordinate, with insertions of two point-like sources at $\sg_1,\sg_2$, the momenta $q_1,q_2$ being their charges. The first electric field is $\Vec{E}^1\equiv(\bar{e}^2_2,-\bar{e}^2_1)$ and the second is $\Vec{E}^2\equiv(-\bar{e}^1_2,\bar{e}^1_1)$. It is worth noting that the gauge condition \eqref{fx_vln_sg_pl} implies that the two fields are the dual of each other, i.e., $E^1_\al=\ep_{\al\bb} E^2_\bb$. The sources originate from the insertion of the operators. The factor $e^{iq_a X^a(\sg)}$ puts a source at the position $\sg$, with charge $q_a$. Having an $n$-point correlation function means $n$ such sources. In the correlation function we integrate over all possible positions for insertion of the sources. 
Going further with the electrostatic analogy, the energy density of the two electric fields is: $|\Vec{E}^1|^2=|\Vec{E}^2|^2=e^{2\Om}$, which is the WS area measure $\sqrt{\bar g}$. The total electrostatic energy in space $\int|\Vec{E}^i|^2$, is thus the WS area $\bar{A}$. 

Point-like sources cause UV-divergences, in any dimension $d\geq 1$, and, in particular, in two-dimensions. They will show up in the energy density $\sqrt{\bar{g}(\sg_i)}$ at the coordinate of any operator $\OO(q_i)$. They will also appear in $\bar{A}$. From the WS perspective, the operator insertion $\OO(q_i)$ causes the WS to develop an infinitely long ``trumpet'' at $\sg_i$, aligned along the momentum direction $q_i$. In string theory, similar insertions lead to a ``trumpet'' which represents an asymptotically scattered closed string. This trumpet causes an infinite area around $\sg_i$. The electrostatic energy $\bar{A}$ may also ``suffer" from IR-divergences on non-compact spaces. For an electric field created by some source distribution, it can happen only in one- or two-dimensions, and it is solely caused by the monopole moment. The dipole moment and higher moments do not cause any IR-divergences. However, momentum conservation ensures that the monopole moment vanishes. Another IR-divergence may be caused by a constant electric field, which is the general homogeneous solution of \eqref{OP_PI_2}. In the electrostatic case, where this homogeneous solution is not allowed, due to boundary conditions. However, the only reasonable demand for the vielbein is that the metric should become flat as $|\sg|\to\infty$. This demand is satisfied by a constant vielbein, and therefore it will be part of the solution. 

The general solution to \eqref{OP_PI_2} is thus composed out of ``source" part and a constant part:
\begin{equation}
    \bar{e} = \bar{e}_s + \bar{e}_c,
\end{equation}
where $\lim_{|\sg|\to\infty} \bar{e}_s=0 \And  \pd_{\al}\bar{e}_c=0$. Evaluation of the $\Om,\phi$ PI, does not require the explicit form of $\bar e_s$. The only important fact is that $\bar e_s$ is unique. In the orthogonal gauge, $\bar e_s,\bar e_c$ take the form $\bar{e}_s = e^{\Om_0+\ep\phi_0}, \bar e_c = e^{\bar\Om+\ep\bar\phi}$, respectively. Therefore, we can write the general solution for the vielbein $\bar{e}$ as:
\begin{equation}\label{OP_PI_3}
    e^{f_{\Om}+\ep f_{\phi}} \equiv e^{\Om_0+\ep\phi_0} + e^{\bar{\Om}+\ep\bar{\phi}}.\footnote{Although it is not obvious that the same diffeomorphism will bring the three $\bar{e},\bar{e}_s,\bar{e}_c$ to the orthogonal form, it will be shown by direct computation that this is indeed the case.}
\end{equation}
The modulus and phase of the summands relates to the modulus and phase of the result (in analogy with addition of complex numbers):
\begin{equation}\label{OP_PI_4}
\begin{split}
    & f_{\Om}(\Om_0,\bar{\Om},\phi_0,\bar{\phi}) \equiv \frac{1}{2}\ln\left(e^{2\Om_0}+e^{2\bar{\Om}}+2e^{\Om_0+\bar{\Om}}\cos(\phi_0-\bar{\phi})\right),
    \\
    & f_{\phi}(\Om_0,\bar{\Om},\phi_0,\bar{\phi}) \equiv \tan^{-1}\left(\frac{e^{\Om_0}\sin(\phi_0)+e^{\bar{\Om}}\sin(\bar{\phi})}{e^{\Om_0}\cos(\phi_0)+e^{\bar{\Om}}\cos(\bar{\phi})}\right).
\end{split}
\end{equation}

The existence of many solutions to the $\de(K_b)$-function constraint, labeled by different $\bar{\Om},\bar{\phi}$, means that we cannot integrate the whole $\Om,\phi$ space immediately. Rather, one should separate the zero eigenvectors. In general, this is done by the following procedure. Suppose one has to evaluate the integral:
\begin{equation}\label{int_split_1}
    \int \prod_{i=1}^n dx_i \prod_{j=1}^m\de(A_{jk}x_k) f(\{x_i\}),
\end{equation}
where $m<n$, and the matrix $A$ is of full rank. Using the inner product $<\bar x,\bar y> = \sum x_i y_i$, where $\bar x = \sum x_i\bar s_i$ with $\bar s_i$ the standard basis, one finds an alternative orthonormal basis, composed of zero modes of $A$ and its orthogonal complement. Let $C$ be the orthogonal matrix transforming between the two bases. We denote by $u_i,\ i=1,\cdots,n-m$ and $v_l,\ l=1,\cdots,m$ the coordinates of the zero subspace and its orthogonal complement, respectively. Then the integral \eqref{int_split_1} becomes:
\begin{equation}
    \int \prod_{i=1}^{n-m} du_i \prod_{l=1}^{m} dv_l \prod_{j=1}^m\de(B_{jk}v_k) \tilde f(\{u_i\},\{v_l\}),
\end{equation}
where $B$ is $m$-by-$m$ matrix, appearing in the product $AC$, the remaining $m$-by-$(n-m)$ part being zero, i.e.: $AC = \begin{pmatrix}
    B_{m \times m} & 0_{m \times (n-m)}
\end{pmatrix}$.
By definition, $B$ is invertible, and so the integral is evaluated easily:
\begin{equation}
    \int \prod_{i=1}^{n-m} du_i \frac{\tilde f(\{u_i\})}{|\det(B)|}.
\end{equation}
We can write:
\begin{equation}
    |\det(B)| = \sqrt{\det(B^{\dagger}B)} = \sqrt{{\det}'(C^{\dagger}A^{\dagger}AC)} = \sqrt{{\det}'(A^{\dagger}A)}.
\end{equation}
The second equality results from the relation: $(AC)^{\dagger}(AC) = \begin{pmatrix}
    B_{m \times m} & 0_{m \times (n-m)} \\
    0_{(n-m) \times m} & 0_{(n-m) \times (n-m)}
\end{pmatrix},$
where ${\det}'$ excludes the zero eigenvalues. Then the final result is:
\begin{equation}\label{int_split_2}
    \int \prod_{i=1}^{n-m} du_i \frac{\tilde f(\{u_i\})}{\sqrt{{\det}'(A^{\dagger}A)}}.
\end{equation}

Returning to the $\Om,\phi$ integral, for a functional $\mathcal{F}$ of $\bar e$:
\begin{equation}\label{OP_PI_5}
    \int \pr{\prod_{\sg=1}^N d\Om(\sg) d\phi(\sg)} \prod_{b=1}^2\prod_{\sg'=1}^{N-1} \de(K_b(\sg'))\mathcal{F}(\Om(\sg),\phi(\sg)) = \int d\bar{\Om}d\bar{\phi} \sqrt{G^{\bar{\Om},\bar{\phi}}} \frac{\mathcal{F}(f_{\Om},f_{\phi})}{\sqrt{{\det}'(Q^{\dagger}Q)}},
\end{equation} 
Here, $Q(f_{\Om},f_{\phi})$ denotes the linearization of the constraints $K_a(\sg)$ around the solution $f_{\Om},f_{\phi}$. It is given by:
\begin{equation}
    Q
    \begin{pmatrix}
        \de\Om \\ \de\phi    
    \end{pmatrix}
    \equiv \ep^{\al\bb}\left(\pd_\al\bar{e}^c_{\bb} + \bar{e}^c_{\bb}\pd_\al \right)
    \begin{pmatrix}
        \de^b_c & \ep^b_c
    \end{pmatrix}
    \begin{pmatrix}
        \de\Om \\ \de\phi    
    \end{pmatrix}.
\end{equation}
As for $\bar x$ mentioned above, the volume element $\displaystyle \prod_{\sg=1}^N d\Om(\sg) d\phi(\sg)$ is induced from the inner product $\displaystyle \sum_\sg (d\Om(\sg)d\Om'(\sg) + d\phi(\sg)d\phi'(\sg))$. This inner product $G$ is used to calculate $\sqrt{G^{\bar{\Om},\bar{\phi}}}$, the volume element in the two dimensional zero mode subspace. It is needed also in the calculation of $Q^{\dagger}$, as well as a similar one $\displaystyle \sum_{a=1}^2 \sum_\sg dY^a(\sg) dY'^a(\sg)$.\footnote{In principle we could have done the whole analysis with the inner products $G_Y$ and $G_s$ for the $Y^a(\sg)$ and $\Om(\sg),\phi(\sg)$ spaces respectively, but the calculations are cleaner in this form.} The following will be devoted to evaluation of $\sqrt{G^{\bar{\Om},\bar{\phi}}}$. The evaluation of $|{\det}'(Q)|$ is postponed until subsection \ref{subsec_PI_vs}, after deriving the explicit solution of \eqref{OP_PI_2}. The calculation of $\sqrt{G^{\bar{\Om},\bar{\phi}}}$ requires finding the infinitesimal vectors: 
\begin{equation}
\begin{split}
    \de U_1 &= \frac{\pd f_\Om}{\pd\bar\Om}d\bar\Om + \frac{\pd f_\phi}{\pd\bar\Om}d\bar\Om, \\
    \de U_2 &= \frac{\pd f_\Om}{\pd\bar\phi}d\bar\phi + \frac{\pd f_\phi}{\pd\bar\phi}d\bar\phi,
\end{split}
\end{equation} 
for each $\bar\Om,\bar\phi$, evaluating the inner product $G^{\bar{\Om},\bar{\phi}}_{ij} = G_s(\de U_i,\de U_j)$ between them, and taking the determinant. A short calculation gives:
\begin{equation}
    \begin{split}
        \de f_\Om &= e^{\bar\Om - f_\Om} \left( \cos\left(f_\phi - \bar\phi\right) d\bar\Om + \sin\left(f_\phi - \bar\phi\right) d\bar\phi  \right)  
        \\
        \de f_\phi &= e^{\bar\Om - f_\Om} \left( -\sin\left(f_\phi - \bar\phi\right) d\bar\Om + \cos\left(f_\phi - \bar\phi\right) d\bar\phi  \right) 
    \end{split}
    \Biggr\} \Rightarrow
    G^{\bar{\Om},\bar{\phi}}_{ij} = |\La|\de_{ij} \int d^2\sg e^{2\bar\Om-2f_\Om},
\end{equation}
and therefore $\sqrt{G^{\bar{\Om},\bar{\phi}}} = \int d^2\sg e^{2\bar\Om-2f_\Om}$. It is easy to verify that the operator $Q$ annihilates the two zero modes:
\begin{equation}\label{eigenvectors_Q}
    e^{\bar\Om - f_\Om}
    \begin{pmatrix}
        \cos\left(f_\phi - \bar\phi\right) \\
        -\sin\left(f_\phi - \bar\phi\right)
    \end{pmatrix}
    ,\ 
    e^{\bar\Om - f_\Om}
    \begin{pmatrix}
        \sin\left(f_\phi - \bar\phi\right) \\
        \cos\left(f_\phi - \bar\phi\right)
    \end{pmatrix}.
\end{equation}
Using \eqref{OP_PI_5} in \eqref{cor_dev_6} gives:
\begin{equation}\label{cor_dev_7}
        \begin{split}
        e^{|\La|\ATS} \de(q_1+q_2) \int &\DD\psi d\bar{\Om}d\bar{\phi} \left(|\La| \int d^2\sg e^{2\bar\Om-2f_\Om}\right) d\bar A(\sg_1)d\bar A(\sg_2)\de(\sg_1) \frac{\left(\det(e^{2f_\Om})\right)^2 J(\bar{e},0)}{|\sqrt{{\det}'(Q^{\dagger}Q)|}} \cdot \\
        &  \qquad O(\sg_1)O(\sg_2) e^{iq_1\cdot \sg_1 + iq_2\cdot \sg_2 - S_0(\psi,\bar{e}) - |\La|\left(\bar{A} + i\int d^2\sg\bar{e}^a_\al I^\al_a\right)},
    \end{split}
\end{equation}
where $\bar{e}$ is evaluated at $f_{\Om},f_{\phi}$. As the $\Om$ and $\phi$ PI's were evaluated, we will use $\Om,\phi$ instead of $f_{\Om},f_{\phi}$. 

% ------------------------------------
\subsection{The Undeformed Fields Path Integral}

The last PI to be evaluated is the $\psi$ PI: 
\begin{equation}\label{P_PI_1}
    \int\DD\psi O(\sg_1)O(\sg_2) e^{-S_0(\psi,e^{2\Om}\de)}.
\end{equation}
This PI is difficult to evaluate for a general QFT, and thus the following is restricted to undeformed theories which are CFTs\footnote{whereas up to here the analysis was valid for a general QFT}. As mentioned in the introduction, in this case, the coupling of the metric to the undeformed theory can be chosen to be such that $S_0(\psi,g)$ has Weyl invariance, i.e., it is invariant under a simultaneous rescaling of the fields and the metric:
\begin{equation}\label{P_PI_2}
    \psi = e^{-\Om\Delta_{\psi}}\tilde\psi \And g = e^{2\Om}\tilde g \qquad \Rightarrow \qquad S_0(\tilde\psi,\tilde g) = S_0(\psi,g).
\end{equation}
Then, for a primary operator $O(\sg)$ with conformal dimension $\Delta_O$, defining transformed fields and operators $\psi = e^{-\Om \Delta_{\psi}} {\tilde{\psi}}$ and $O= e^{-\Om \Delta_O} {\tilde O}$, we have:
\begin{equation}\label{P_PI_3}
    \begin{split}
         & \int\DD\psi O(\sg_1)O(\sg_2) e^{-S_0\left(\psi,e^{2\Om}\de\right)} = \int\DD\psi e^{-(\Om(\sg_1)+\Om(\sg_2))\Delta_O}{\tilde O}(\sg_1){\tilde O}(\sg_2) e^{-S_0\left(e^{-\Om\Delta_{\psi}}\tilde{\psi},e^{2\Om}\de\right)} = \\
         & e^{-(\Om(\sg_1)+\Om(\sg_2))\Delta_O +W(\sg_1,\sg_2)} \int\DD\tilde{\psi} {\tilde O}(\sg_1){\tilde O}(\sg_2)e^{-S_0(\tilde{\psi},\de)} =
         e^{-(\Om(\sg_1)+\Om(\sg_2))\Delta_O +W(\sg_1,\sg_2)}Z_0 \frac{C}{|\sg_{21}|^{2\Delta_O}},
    \end{split}
\end{equation}
where $\sg_{21} \equiv \sg_2 -\sg_1$, $C$ is the constant in the two-point correlator of the operator $O$ in the undeformed theory, and $Z_0$ is its partition function. The factor $e^W$ is the Weyl anomaly coming from the change of variables $\psi \to \tilde{\psi}$, i.e. $\DD\psi = e^W \DD\tilde\psi$. It takes the form:
\begin{equation}\label{Weyl_1}
    W = \frac{c}{24\pi}\int d^2\sg \de^{\al\bb}\pd_\al\Om\pd_\bb\Om,
\end{equation}
were $c$ is the central charge of the undeformed theory.

Using \eqref{P_PI_3} inside \eqref{cor_dev_7} gives finally: 
\begin{equation}\label{cor_dev_8}
\begin{split}
    Z_0 e^{|\La|\ATS} \de(q_1+q_2) \int d\bar\Om d\bar\phi &\left(|\La| \int d^2\sg e^{2\bar\Om-2f_\Om}\right) d^2\sg_1 d^2\sg_2 \de(\sg_1) \frac{\left(\det(e^{2f_\Om})\right)^2 J(\bar{e},0)}{|\sqrt{{\det}'(Q^{\dagger}Q)|}} \cdot
    \\ 
    & \frac{C}{|\sg_{21}|^{2\Delta}} e^{(2-\Delta)(\Om(\sg_1)+\Om(\sg_2))}e^{iq_1\cdot \sg_1+iq_2\cdot \sg_2 + W - |\La|\left(\bar{A} + i\int d^2\sg\bar{e}^a_\al I^\al_a\right)},    
\end{split}
\end{equation}
where the area elements $d\bar A(\sg_1),d\bar A(\sg_2)$ were written explicitly, and $\Delta\equiv\Delta_O$.

% ------------------------------------
\subsection{Vielbein Solution}\label{subsec_PI_vs}

% ------------
\subsubsection{General Case}

It remains to find the non-homogeneous solution $\bar{e}_s$ to \eqref{OP_PI_1}, which vanishes at infinity. Transforming \eqref{fx_vln_sg_pl} from Cartesian coordinates to complex coordinates $w,\bar w$, the gauge-fixed metric and vielbein take the form:
\begin{equation}\label{fx_mtrc_w}
    [\bar g_{\al\bb}]_{w}=e^{2\Om(\sg)}
    \begin{pmatrix}
        0           & \frac{1}{2}  \\ 
        \frac{1}{2} & 0
    \end{pmatrix},    
\end{equation}
\begin{equation}\label{fx_vln_w}
    [\bar{e}^a_{\al}]_{w} = e^{\Om(\sg)}e^{\ep\phi(\sg)}
    \begin{pmatrix}
        \frac{1}{2}  & \frac{1}{2}  \\ 
        -\frac{i}{2} & \frac{i}{2}
    \end{pmatrix}    
    =
    \frac{1}{2}e^{\Om}
    \begin{pmatrix}
        e^{-i\phi}  & e^{i\phi}  \\ 
        -ie^{-i\phi} & ie^{i\phi}
    \end{pmatrix}
    \equiv
    \frac{1}{2}
    \begin{pmatrix}
        \bar{z} & z \\
        -i\bar{z} & iz
    \end{pmatrix},
\end{equation}
where $z\equiv e^{\Om+i\phi}$. In these coordinates, \eqref{OP_PI_1} takes the form:
\begin{equation}\label{vs_3}
    -\frac{i}{2}
    \begin{pmatrix}
        i(\pd z+\bpd\bar{z}) \\
        -\pd z+\bpd\bar{z}
    \end{pmatrix}
    =
    \begin{pmatrix}
        Q_1 \\
        Q_2
    \end{pmatrix}
    2\pi(-\de(w-w_1)+\de(w-w_2)), 
\end{equation}
or
\begin{equation}\label{vs_4}
    \begin{pmatrix}
        \pd z \\ \bpd\bar{z}
    \end{pmatrix}
    =
    -\begin{pmatrix}
        \bar{Q} \\ Q
    \end{pmatrix}
    2\pi(\de(w-w_1)-\de(w-w_2)), 
\end{equation}
where:
\begin{equation}
\begin{split}
    & Q_a \equiv \frac{(q_1)_a}{2\pi|\La|} = -\frac{(q_2)_a}{2\pi|\La|}, \\
    & Q \equiv Q_1 + iQ_2,\ \bar{Q} \equiv Q_1 - iQ_2.\footnote{We hope that the momentum $Q$ defined here will not be confused with the operator $Q$ used before, which we defined following the notation of \cite{Dubovsky:2018bmo}.}
\end{split}
\end{equation}
The two equations are complex conjugates of each other, as should happen in complex coordinates. \eqref{vs_4} is a linear equation which is easily solved. In the region $\mathbb{C}\setminus \{w_1,w_2\}$, the function $z$ satisfies $\pd z=0$. Therefore, it is anti-holomorphic, and has a Laurent-series expansion. The boundary condition at infinity, i.e. $z \to \text{const}$, excludes all the positive powers, and the $\de$-function singularities at $w_1,w_2$ allow only simple poles. Hence, the general solution is:
\begin{equation}\label{vs_6}
    z=-\bar{Q}\left( \frac{1}{\bar{w}-\bar{w}_1}-\frac{1}{\bar{w}-\bar{w}_2} \right) + e^{\bar{\Om}+i\bar{\phi}},
\end{equation}
where $\bar{\Om},\bar{\phi}$ are constants.

The divergences in $[\bar g_{\al\bb}]_\sg = |z|^2\de_{\al\bb}$ at $w=w_{1,2}$, introduce UV-divergences in the correlation function. Examining \eqref{cor_dev_8}, except for the determinants, they appear in the following factors. 
\newline The first are the area elements $\sqrt{\bar{g}(\sg_1)},\sqrt{\bar{g}(\sg_2)}$, which are divergent. 
\newline The second divergence comes through the WS area $\bar{A}$. For this we need the area element (in Cartesian coordinates):
\begin{equation}\label{vs_7}
    e^{2\Om} = |Q|^2 \frac{|w_{21}|^2}{|w-w_1|^2|w-w_2|^2} - 2\Re\left( \frac{Qw_{21}e^{i\bar{\phi}}}{(w-w_1)(w-w_2)} \right)e^{\bar{\Om}} + e^{2\bar{\Om}},
\end{equation}
where $w_{21}\equiv w_2-w_1$. In our conventions $|w|=|\sg|$. Upon integration over $w$ (with any UV-regularization that respects translation and rotation symmetries, in particular point-splitting regularization), the middle term in \eqref{vs_7} vanishes. Integration of the first term with a point-splitting regularization yields a logarithmic divergence:
\begin{equation}
    \int \frac{d^2w}{2}\frac{|w_{21}|^2}{|w-w_1|^2|w-w_2|^2} = \int \frac{d^2w}{2}\frac{|w_{21}|^2}{|w|^2|w-w_{21}|^2} = 4\pi \lim_{\ep\rightarrow 0^+} \ln\left(\frac{|w_{21}|}{2\ep}\right) \rightarrow \infty,
\end{equation}
($d^2\sg = d^2w/2$, see \eqref{compcoor}).
Integration of the last term gives an IR-divergent term $e^{2\bar{\Om}}\ATS$. Gathering both terms gives the naive WS area:
\begin{equation}\label{vs_8}
    \bar{A} = 4\pi|Q|^2\lim_{\ep\rightarrow 0^+} \ln\left(\frac{|w_{21}|}{2\ep}\right) + e^{2\bar{\Om}} \ATS.
\end{equation}
We got an IR-divergent term only from the constant part $e^{2\bar{\Om}}$, not from the ``charges", in accordance with the discussion in subsection \ref{subsec_PI_OP}.
\newline The third divergence occurs in the Weyl factor $W$ in \eqref{Weyl_1}. It will be discussed in the next section. 
\newline The fourth and last divergence appears in the zero modes volume element $\int d^2\sg e^{-2\Om}$. In the next subsection this factor will cancel with a similar factor from the determinant of $Q^{\dagger}Q$.

One last term in \eqref{cor_dev_8} needs to be evaluated, which is $\int d^2\sg \bar{e}^a_\al I^\al_a$. The integrand is the trace of the vielbein, which is:
\begin{equation}\label{vs_9}
    \bar{e}^a_\al I^\al_a = 2e^{\Om}\cos(\phi) = -2\Re\left( \frac{Qw_{21}}{(w-w_1)(w-w_2)} \right) + 2e^{\bar{\Om}}\cos(\bar{\phi}).
\end{equation}
The integral of the first term vanishes, for the same reasoning regarding the middle term of \eqref{vs_7}. The second term contributes $2e^{\bar{\Om}}\cos(\bar{\phi}) \ATS$. Using \eqref{vs_8} and \eqref{vs_9}, the last part of \eqref{cor_dev_8} is:
\begin{equation}\label{vs_10}
    e^{-|\La|\left(\bar{A} + i\int d^2\sg\bar{e}^a_\al I^\al_a\right)} = \lim_{\ep\rightarrow 0^+}\left(\frac{|w_{21}|}{2\ep}\right)^{-4\pi|\La||Q|^2} e^{-|\La|\ATS\left(e^{2\bar{\Om}} + 2ie^{\bar{\Om}}\cos(\bar{\phi}) \right)}.
\end{equation}

% ------------
\subsubsection{Determinants I}

Having the explicit form of the vielbein, $\sqrt{{\det}'(Q^{\dagger}Q)}$ can be calculated, following the same steps as in subsection \ref{subsec_FP_gft1} for the operator $P$, with the inner products mentioned at the end of subsection \ref{subsec_PI_OP}. For 
$Q^{\dagger}$ one finds:
\begin{equation}
    Q^{\dagger}\de K_b \equiv 
    -
    \begin{pmatrix}
        \de^b_c \\
        \ep^b_c
    \end{pmatrix}    \ep^{\al\bb}\bar{e}^c_{\bb}\pd_{\al}\de K_b.
\end{equation}
Composing $Q^{\dagger}$ with $Q$ gives:
\begin{equation}
\begin{split}
    Q^{\dagger}Q &= 
    \begin{pmatrix}
        \ A & B\ \\
        -B & A\ 
    \end{pmatrix}, 
    \\
    A &\equiv -e^{2\Om} \de^{\al\bb} \left( \pd_\al\pd_\bb + \pd_\al\Om\pd_\bb + \pd_\bb\Om\pd_\al + \pd_\al\Om\pd_\bb\Om - \pd_\al\phi\pd_\bb\phi +\pd_\al\pd_\bb\Om \right) + 2e^{2\Om}\ep^{\al\bb}\pd_\al\Om\pd_\bb\phi,
    \\
    B &\equiv e^{2\Om} \de^{\al\bb} \left(\pd_\al\phi\pd_\bb + \pd_\bb\phi\pd_\al + \pd_\al\phi\pd_\bb\Om + \pd_\bb\phi\pd_\al\Om + \pd_\al\pd_\bb\phi \right).
\end{split}
\end{equation}
One can verify that $A^{\dagger}=A,\ B^{\dagger}=-B$, as required from the hermiticity of $Q^{\dagger}Q$. It is easier to use a (block) diagonal form of $Q^{\dagger}Q$:
\begin{equation}
\begin{split}
    &
    \begin{pmatrix}
        \ A & B\ \\
        -B & A\ 
    \end{pmatrix} 
    =
    U
    \begin{pmatrix}
        A+iB & 0 \\
        0 & A-iB \ 
    \end{pmatrix}
    U^\dagger ,
    \\
    & U \equiv \frac{1}{\sqrt{2}}
    \begin{pmatrix}
        I & I\ \\
        iI & -iI\ 
    \end{pmatrix}.
\end{split}
\end{equation}
The explicit form of the operators $A\pm iB$ can be simplified using $z$:
\begin{equation}
\begin{split}
    A+iB &= -4e^{2\Om}(\pd + \pd\ln(\bar z))(\bpd + \bpd\ln(\bar z)), 
    \\
    A-iB &= -4e^{2\Om}(\bpd + \bpd\ln(z))(\pd + \pd\ln(z)).
\end{split} 
\end{equation}
It is easy to verify that $A\pm iB$ each has only one zero eigenvector, which is $1/\bar z,\ 1/z$, respectively.\footnote{These eigenvectors are the suitable linear combinations of the eigenvectors in \eqref{eigenvectors_Q}.} Finally, the determinant is:
\begin{equation}\label{det_Q_1}
    \sqrt{{\det}'(Q^{\dagger}Q)} = \left|{\det}\pr{-e^{2\Om}(\pd +\pd\ln(\bar z))(\bpd + \bpd\ln(\bar z))}\right|,
\end{equation}
where the constant ${\det}'(4)$ was omitted.\footnote{Naively $\pd\ln(z) = \bpd\ln(\bar z) = 0$, since the derivative gives a sum of $\de$-functions, each one multiplied by its argument. However, by smearing of the $\de$-function, which is physically more reasonable, they will not vanish.}

% ------------
\subsubsection{Determinants II}

Using \eqref{det_dec_1} and \eqref{det_Q_1}, the determinant ratio in \eqref{cor_dev_7} can be simplified:
\begin{equation}\label{det_dec_2}
    \frac{\left(\det(e^{2\Om})\right)^2 J(\bar{e},0)}{\sqrt{{\det}'(Q^{\dagger}Q)}} = \frac{\left|\det(e^{2\Om})\right|\left|\det(-\pd\bpd-4\pd\Om\bpd+\de(w))\right|}{\left|{\det}'\pr{-e^{2\Om}(\pd +\pd\ln(\bar z))(\bpd + \bpd\ln(\bar z))}\right|}.
\end{equation}
We would like to cancel $\det(e^{2\Om})$ from the numerator and denominator, but because of the zero mode exclusion in the denominator, we can not naively split ${\det}'\left(e^{2\Om}\square \right) \rightarrow {\det} \left(e^{2\Om}\right) {\det}'(\square)$, where $\square \equiv (\pd +\pd\ln(\bar z))(\bpd + \bpd\ln(\bar z))$. In order to ``resolve" the zero modes, the following regularization will be made. The operator $D\equiv e^{2\Om}\square$ is modified to $D_\ep \equiv e^{2\Om}(\square + \ep)$, where $\ep$ is a small number that will be taken to zero soon. The eigenvalues $\lambda^\ep_n$ of $D_\ep$, are slightly shifted from $\lambda_n$ of $D$: $\lambda^\ep_n \approx \lambda_n + \ep \al_n$. The eigenvalue $\lambda_0 = 0$ was excluded in ${\det}'(D)$, and therefore $\lambda^\ep_0 \approx 0 + \ep\al_0$ will be excluded in the regularized version. Using these definitions we evaluate:
\begin{equation}
\begin{split}
    {\det}'(D_\ep) &= \prod_{n\neq 0}\lambda^\ep_n = \frac{\prod_{n}\lambda^\ep_n}{\lambda^\ep_0} = \frac{\det(D_\ep)}{\ep\al_0} = \frac{\det(e^{-2\Om})\det(\square + \ep)}{\ep\al_0} = \det(e^{2\Om}) \frac{{\det}'(\square + \ep)\ep}{\ep\al_0}  \\
    &\xrightarrow[{\ep \rightarrow 0}]{} {\det}'(D) =  \frac{1}{\al_0}\det(e^{2\Om}){\det}'(\square).
\end{split}
\end{equation}
Hence, splitting the determinant is correct up to the finite factor $\al_0^{-1}$, which still needs to be calculated. For this aim, writing $f^\ep_0 \approx \frac{1}{\bar z} + \ep f_0$, one needs to solve the following eigenvalue equation:
\begin{equation}\label{det_Q_2}
e^{2\Om}\left(\square + \ep\right)f^\ep_0 = \lambda^\ep_0 f^\ep_0 \Rightarrow \pd\bpd(\bar z f_0) = \frac{\al_0}{e^{2\Om}} - 1.
\end{equation}
Boundary conditions determine $f_0$ and $\al_0$ completely. Considering the torus IR-regularization, $f^\ep_0$, and therefore also $f_0$, can at most can be constant as $|\sg| \rightarrow \infty$. Then, the norm of $f^\ep_0$, $\int d^2\sg |f^\ep_0|^2$, has an IR-divergence of $\ATS$, which is finite in the torus.  $\bar z \to \text{const}$ at infinity, therefore $\int d^2\sg \pd\bpd(\bar z f_0) = 0$. Then, integration of \eqref{det_Q_2} gives: $\al_0 = \ATS / \left( \int d^2\sg e^{-2\Om} \right)$. In conclusion, \eqref{det_dec_2} becomes:
\begin{equation}\label{det_dec_3} 
    \frac{\left|\det(-\pd\bpd-4\pd\Om\bpd+\de(w))\right|}{\left|{\det}'\pr{-(\pd +\pd\ln(\bar z))(\bpd + \bpd\ln(\bar z))}\right|} \cdot \frac{\ATS}{\int d^2\sg e^{-2\Om}} \equiv D(z,0) \cdot \frac{\ATS}{\int d^2\sg e^{-2\Om}} .
\end{equation}
Substituting this result in \eqref{cor_dev_8} gives:
\begin{equation}\label{cor_dev_8a}
\begin{split}
    Z_0 |\La|\ATS e^{|\La|\ATS} \de(q_1+q_2) \int & d\bar{\Om}d\bar{\phi} d^2\sg_1 d^2\sg_2 \de(\sg_1) D(z,0) \cdot
    \\ 
    & \frac{C}{|\sg_{21}|^{2\Delta}} e^{(2-\Delta)(\Om(\sg_1)+\Om(\sg_2))}e^{iq_1\cdot \sg_1+iq_2\cdot \sg_2 + 2\bar\Om + W - |\La|\left(\bar{A} + i\int d^2\sg\bar{e}^a_\al I^\al_a\right)}.
\end{split}
\end{equation}

At the end of this subsection, we note one consequence of the solution in \eqref{vs_7}, which is related to the determinants appearing in \eqref{det_dec_3}. In the large momentum limit, i.e., when $|Q|$ is much larger than any length scale in the problem, we can approximate:
\begin{equation}
    \bar z \approx \frac{Qw_{21}}{(w-w_1)(w-w_2)} \Rightarrow \ln \bar z = \Om + i\phi \approx \ln(Q) + \ln\left(\frac{w_{21}}{(w-w_1)(w-w_2)}\right).
\end{equation}
In this limit, $\pd\ln(\bar z),\bpd\ln(\bar z)$ (and hence $\pd\Om$ and $\bpd\Om$) do not depend on the momenta, and therefore also the two determinants in \eqref{det_dec_3}.

% ------------
\subsubsection{Degenerate Case}

The limit where $|\La| \rightarrow \infty$ (no $\TT$-deformation), corresponds to the degenerate case $|Q|=0$. Then, the source part in the vielbein vanishes ($e^{\Om_s}=0$), and 
$\Om(\sg)=\bar{\Om},\ \phi(\sg)=\bar{\phi}$. The WS is flat, the area element is $dA(\sg)=e^{2\bar{\Om}}d^2\sg$, and the WS area \eqref{vs_8} reduces to $\bar{A}=e^{2\bar{\Om}}\ATS$. The determinant ratio \eqref{det_dec_3} becomes:
\begin{equation}\label{det_dec_4}
     \frac{\left|\det(-\pd\bpd + \de(w))\right|}{\left|{\det}'(-\pd\bpd)\right|} \equiv D(0),
\end{equation}
and can be pulled out of the integral over $\sg_{1,2}$. $W$ also vanishes. No UV-divergences appear and there is no need for UV-regularization. Therefore, \eqref{cor_dev_8a} can be written as:
\begin{equation}
    \begin{split}
        Z_0|\La|\ATS e^{|\La|\ATS} \de(q_1+q_2) \int & d\bar{\Om}d\bar{\phi} d^2\sg_1 d^2\sg_2 \de(\sg_1) D(0) \cdot
        \\
        & \frac{C}{|\sg_{21}|^{2\Delta}} e^{(3-\Delta)(2\bar{\Om})}e^{iq_1\cdot \sg_1+iq_2\cdot \sg_2 - |\La|\ATS\left(e^{2\bar{\Om}} + ie^{\bar{\Om}}\cos(\bar{\phi})\right)},
    \end{split}
\end{equation}
which can be organized as follows:
\begin{equation}
    \begin{split}
        Z_0|\La|\ATS  e^{|\La|\ATS} D(0) & \int d\bar{\Om}d\bar{\phi} e^{2(3 - \Delta)\bar{\Om}} e^{-|\La|\ATS\left(e^{2\bar{\Om}} + ie^{\bar{\Om}}\cos(\bar{\phi})\right)} \cdot 
        \\
        \de(q_1+q_2) & \int d^2\sg_1 d^2\sg_2 \de(\sg_1)
        \frac{C}{|\sg_{21}|^{2\Delta}} e^{iq_1\cdot \sg_1+iq_2\cdot \sg_2}.
    \end{split}
\end{equation}
The first line describes an IR-divergent factor and does not depend on the momentum. Part of it cancels upon dividing by $Z_{\TT}$, and the rest can be absorbed by an appropriate normalization of the PI. The second line is the Fourier transform of the undeformed correlator, as expected.

% ------------------------------------
\subsection{$n$-Point Correlator} \label{subsec_npoint}

The calculation of the $n$-point correlation function $C(q_1,\cdots,q_n)$, with operators of dimensions $\Delta_i$, follows the same steps as above. The analogue of \eqref{cor_dev_8a} reads:
\begin{equation}\label{cor_dev_n_1}
\begin{split}
    Z_0 |\La|\ATS e^{|\La|\ATS} &\de\left(\sn q_i\right) \int \pn d^2\sg_i \de(\sg_1) \int d\bar{\Om}d\bar{\phi} D(z,0) \cdot
    \\
    & \left\langle \pn O(\sg_i) \right\rangle \left(\pn e^{(2-\Delta_i)\Om(\sg_i)}\right) e^{ i\sum_{i=1}^n q_i\cdot \sg_i + 2\bar\Om + W - |\La|\left(\bar{A} + i\int d^2\sg\bar{e}^a_\al I^\al_a\right)}.
    \footnotemark
\end{split}
\end{equation}
\footnotetext{As for the two-point correlator, in \eqref{cor_dev_n_1} the integral over $\sg_1$ was not evaluated, in order to maintain the symmetry between all the operators' positions $\sg_i$. Using the $\de(\sg_1)$ its value could be fixed to zero, so the remaining $\sg_i$ ($i \geq 2$) are essentially $\sg_{i1}$.}
In \eqref{cor_dev_n_1}, $\left\langle \pn O(\sg_i) \right\rangle \equiv C^0(\{\sg_i\})$ is the undeformed correlator, which in the case of the two-point correlator was simply $\frac{C}{|\sg_{21}|^{2\Delta}}$. \eqref{vs_4} will be modified to:
\begin{equation}\label{vs_n_1}
    \pd z = -\sn 2\pi \bar{Q}^i \de(w-w_i),
\end{equation}
where, as for the two-point function:
\begin{equation}\label{def_Q}
    Q^i_a \equiv \frac{(q_i)_a}{2\pi|\La|}, \ Q^i \equiv Q^i_1+iQ^i_2.
\end{equation}
The solution is:
\begin{equation}\label{vs_n_2}
    z = -\sn \frac{\bar{Q}^i}{\bar{w}-\bar{w}_i} + e^{\bar{\Om}+i\bar{\phi}}.
\end{equation}
The area element in Cartesian coordinates, $|z|^2=e^{2\Om}$, reads:
\begin{equation}\label{vs_n_3}
    e^{2\Om} = \sn \frac{|Q^i|^2}{|w-w_i|^2} + 2\Re\left(\sum_{i<j}\frac{Q^i\bar{Q}^j}{(w-w_i)(\bar{w}-\bar{w}_j)}\right) - 2\Re\left(\sn \frac{Q^i e^{i\bar{\phi}}}{w-w_i}\right)e^{\bar{\Om}}  + e^{2\bar{\Om}}.
\end{equation}
For the WS area $\bar A$, \eqref{vs_n_3} should be integrated over all coordinate space (using $d^2\sg = \frac{d^2w}{2}$). As a consequence of  rotation and translation symmetries of the coordinate space, the integration $\int d^2w$ of the third summand in \eqref{vs_n_3} vanishes. The integration over the last term in \eqref{vs_n_3} gives an IR-divergent contribution $e^{2\bar\Om}\ATS$. As for the two-point correlator, this divergence arises because of the constant part in the vielbein, not from the source part. The integral over the first summand in \eqref{vs_n_3} is simple:
\begin{equation}\label{vs_n_4}
    \int \frac{d^2w}{2}\frac{1}{|w-w_i|^2} = 2\pi \ln(r)\Big|^{\infty}_0,
\end{equation}
which is IR and UV-divergent. It describes the self energy of a point-like charge, which has both divergences in two-dimensions. Finally, the integral over the second summand in \eqref{vs_n_3}, when evaluated using the Cartesian $\sg$ coordinates, splits into two possible contractions of the TS and WS indices:
\begin{equation}
    \frac{Q^i_a Q^j_b(\sg-\sg_i)^{\al}(\sg-\sg_j)^{\bb}}{|\sg-\sg_i|^2|\sg-\sg_j|^2}(\de^{ab}\de_{\al\bb}+\ep^{ab}\ep_{\al\bb}).
\end{equation}
The integral over the $\ep$-contractions vanishes. Using translational symmetry, the remaining $\de$-contraction term depends only on $\sg_{ij}\equiv\sg_i-\sg_j$. It can be evaluated to give:
\begin{equation}\label{vs_n_5}
    \int d^2\sg \frac{\de_{\al\bb}\sg^{\al}(\sg+\sg_{ij})^{\bb}}{|\sg|^2|\sg+\sg_{ij}|^2} = 2\pi \ln(r)\Big|^\infty_{\frac{|\sg_{ij}|}{2}}.
\end{equation}
This integral is also IR-divergent. Its finite part describes the potential energy between the two operators $O(\sg_i),O(\sg_j)$ (two ``electric charges") with distance $|\sg_{ij}|$ between them. Adding the appropriate coefficients in \eqref{vs_n_3} of the two IR-divergences we found in \eqref{vs_n_4} and \eqref{vs_n_5}, the $2\pi \ln(\infty)$ factor is multiplied by:
\begin{equation}
     \sn |Q^i|^2 + 2\sum_{i<j}\de^{ab}Q^i_a Q^j_b = \sum_{a=1}^2\left(\sn Q^i_a \right)^2,
\end{equation}
which vanishes due to momentum conservation. Hence, it is manifest that the lack of a ``monopole moment", prevents an IR-divergence from the source part. The UV-divergence in \eqref{vs_n_4} results from the divergent ``electric field" near the ``point-like sources"\footnote{as it appeared in \eqref{vs_8} for the two-point correlator.}. It is the ``self energy" of the interaction of the source with itself, producing the infinite energy $\ln(|\sg_i-\sg_i|)$. It will remain for now with a limit $\ep \rightarrow 0$:  
\begin{equation}\label{vs_n_6}
    \bar{A} = -2\pi\sn \lim_{\ep\rightarrow 0^+} \ln(\ep)|Q^i|^2 - 2\pi\cdot2\sum_{i<j} \ln\left({\frac{|\sg_{ij}|}{2}}\right)\de^{ab}Q^i_aQ^j_b + e^{2\bar{\Om}}\ATS.
\end{equation} 
This agrees with the two-point correlator \eqref{vs_8} for the particular case of $n=2$.
 
% ====================================
\section{Regularization, Renormalization and the Final Result}\label{sec_reg}

In this section, the IR and UV-divergences mentioned in the last section will be treated, and the large momentum limit of the $n$-point correlator will be derived.

% ------------------------------------
\subsection{Infra-Red Divergences}\label{subsec_reg_IR}

The IR-divergences will be discussed first. A specific regularization we have in mind is the torus, but we will not discuss it in detail, in particular because in a non-local theory it is not completely obvious that the large volume limit of the torus reproduces the plane. This deserves further investigation.

The IR-divergence in \eqref{cor_dev_8} shows up in the limit $|\La|\ATS\to\infty$. This quantity appears there with the variables $\bar{\Om},\bar{\phi}$ as in \eqref{vs_10}, so it leads to
a saddle point in the $\bar{\Om},\bar{\phi}$ integration. The saddle point of
\begin{equation}\label{IR_1}
    e^{2\bar{\Om}} + 2ie^{\bar{\Om}}\cos(\bar{\phi})
\end{equation}
is easily found to be at $e^{\bar{\Om}}=i$, $\bar{\phi}=\pi$ or at $e^{\bar{\Om}}=-i$, $\bar{\phi}=0$\footnote{Recall that the contour of integration for $e^{\bar{\Om}}$ was Wick-rotated, so these values actually correspond to a real vielbein in the original variables.}. Using the values of $\bar{\Om},\bar{\phi}$ in the saddle, the WS area element and the total WS area in \eqref{vs_7},\eqref{vs_8} are given by:
\begin{equation}\label{IR_2}
    \begin{split}
        e^{2\Om} &= |Q|^2 \frac{|w_{21}|^2}{|w-w_1|^2|w-w_2|^2} + 2i\Re\left( \frac{Qw_{12}}{(w-w_1)(w-w_2)} \right) - 1, \\
        \bar{A} &= 4\pi|Q|^2\lim_{\ep\rightarrow 0^+} \ln\left(\frac{|w_{21}|}{2\ep}\right) - \ATS.   
    \end{split}
\end{equation}
The minus signs seem odd, but they result from the analytic continuation made for the vielbein. That is, the ``original" WS element area and total WS area are $(-i)^2$ times the expressions in \eqref{IR_2}.\footnote{Still, it is somewhat confusing. It also appears in two other possibilities for analytic continuations to make the
$Y$ PI well defined: to Wick-rotate the momentum or the deformation parameter. In these possibilities, they should be Wick-rotated back, since they are variables of the correlator, unlike the vielbein which is an integration variable.}. 

The value of the exponent in \eqref{vs_10} at the saddle found above is $e^{-|\La|\ATS}$, which cancels an opposite factor appearing in \eqref{cor_dev_8}. The Gaussian integral around the saddle yields $\frac{\pi}{|\La|\ATS}$, which again cancels another such factor\footnote{In any case, all of these IR-divergent factors do not depend on the momenta.}. 
After the integration over $\bar{\Om},\bar{\phi}$ using the saddle point approximation, we can rewrite \eqref{cor_dev_8a} as:
\begin{equation}\label{cor_dev_9}
        \begin{split}
        Z_0 \pi\de(q_1+q_2) 
        & \int d^2\sg_1 d^2\sg_2 \de(\sg_1) e^{iq_1\cdot \sg_1+iq_2\cdot \sg_2+W}\frac{C}{|\sg_{21}|^{2\Delta}} \cdot
        \\ 
        & \lim_{\ep\rightarrow 0^+} D(z,0) \left(\frac{|\sg_{21}|}{2\ep}\right)^{-4\pi|\La||Q|^2} e^{(2-\Delta)(\Om(\sg_1)+\Om(\sg_2))},
    \end{split}
\end{equation}
where we also used \eqref{vs_10} (and omitted an overall minus from evaluation of $e^{2\bar\Om}$).

% ------------------------------------
\subsection{Ultra-Violet Divergences}\label{subsec_reg_UV}

There are three UV-divergences in \eqref{cor_dev_9}. The first appears in the factors $\sqrt{\bar{g}(\sg_i)}^{\frac{2-\Delta}{2}}$. These are clearly related to the operator insertions at $\sg_i$. Hence, they should be treated by regularization of the operator $\OO$ defined in \eqref{op_def_mom}. Point-splitting regularization will be used, i.e., instead of evaluating $\sqrt{\bar g(\sg)}$ at $\sg_i$ where it is divergent, it will be evaluated at some nearby point $\sg_i^\al+\ep^\al$. The same regularization will be applied to the other divergences. Some direction of $\ep^\al$ should be chosen, and the simplest choice is to take an average over all directions with the same distance from $\sg_i$, i.e.:
\begin{equation}\label{UV_1}
    \ep^\al \equiv \ep (\cos(\al),\sin(\al)) \ ,\  \sqrt{\bar{g}(\sg_i)} \rightarrow \frac{1}{2\pi}\int_0^{2\pi} d\al \sqrt{\bar{g}(\sg_i+\ep^\al)}.
\end{equation}
Pictorially, in the electrostatic analogue of subsection \ref{subsec_PI_OP}, a small circle of radius $\ep$ around each point-like charge is removed. For $\sqrt{\bar g(\sg)}$ the integration of all directions $\al$ cancels the second term in \eqref{IR_2}, and the area element is:
\begin{equation}\label{UV_2}
    \sqrt{\bar{g}(\sg_{i})} \rightarrow \frac{|Q|^2}{\ep^2} - 1.
\end{equation}
Typically, $\ep$ is taken to be much smaller than any other length scale in the theory, in particular, $|Q|/\ep>>1$, and so we can ignore the last term in \eqref{UV_2}. Consequently, the expressions in \eqref{cor_dev_9} can be written as:
\begin{equation}\label{UV_3}
    e^{(2-\Delta)\Om(\sg_1)} = e^{(2-\Delta)\Om(\sg_2)} = \left(\frac{|Q|}{\ep}\right)^{2-\Delta}.
\end{equation}

The same regularization makes $e^{|\La|\bar{A}}$ finite. Practically, it means leaving $\ep$ finite in \eqref{vs_8}. In fact, every operator insertion gives its own divergence in $\bar A$, and hence by using a renormalized operator the UV-divergences of $\bar A$ will be removed completely. To show this, it is convenient to consider the $n$-point correlator.\footnote{The case of the two-point correlator is misleading for two reasons. First, momentum conservation allows one to consider only one momentum out of the two, and write \eqref{vs_6} with a common factor of $Q$ outside the singular part. Second, in the two-point correlator one considers operators $O$ with the same conformal dimension $\Delta$ (otherwise the correlator vanishes), whereas in general, we may consider operators $O_i$ with different conformal dimensions $\Delta_i$.} Using \eqref{vs_n_3}, the analogue of \eqref{UV_2} for the $n$-point correlation function is:
\begin{equation}\label{UV_4}
    \sqrt{\bar{g}(\sg_i)} \rightarrow \frac{|Q^i|^2}{\ep^2} \Rightarrow
    e^{(2-\Delta_i)\Om(\sg_i)} \rightarrow \left(\frac{|Q^i|}{\ep}\right)^{2-\Delta_i}.
\end{equation}
As for the WS area, the point-splitting regularization prevents the divergence of ``self energy", setting the ``self distance'' to $\ep$. It means leaving $\ep$ finite in \eqref{vs_n_6}. Upon the exponentiation $e^{-|\La|\bar{A}}$ in \eqref{cor_dev_n_1}, the regulator-dependent sum in $\bar A$ can be expressed as:
\begin{equation}\label{UV_5}
    e^{|\La| 2\pi \sum_{i=1}^n \ln(\ep)|Q^i|^2} \rightarrow \prod\limits_{i=1}^n \ep^{2\pi|\La||Q^i|^2},
\end{equation}
which factorizes between the operators. 

The last UV-divergent quantity to be evaluated is the Weyl factor $e^W$. Using complex coordinates, \eqref{Weyl_1} is:
\begin{equation}\label{Weyl_2}
    W = \frac{c}{12\pi}\int d^2w \pd\Om \bpd\Om.
\end{equation}
With $\Om = \frac{1}{2}(\ln(z) + \ln(\bar z))$, evaluation of \eqref{Weyl_2} yields three terms. The first is (we omit a factor $\frac{c}{48\pi}$ common to all three terms):
\begin{equation}
    \int d^2w \frac{\bpd \bar z \pd \bar z}{z^2} + c.c. = -2\pi\sn Q^i\frac{\pd \bar z(w_i)}{z^2(w_i)} + c.c. = -\sn Q^i \int_0^{2\pi} d\al \frac{Q^i/(\ep e^{i\al})^2}{(Q^i)^2/(\ep e^{i\al})^2} + c.c. = -4\pi n.
\end{equation}
The second is:
\begin{equation}
    \int d^2w \frac{\pd z \bpd \bar z}{|z|^2} = \int d^2w \frac{4\pi^2}{|z|^2}\sum_{i,j=1}^n Q^i \bar Q^j\de(w-w_i)\de(w-w_j).
\end{equation}
When $i \neq j$ the integral clearly vanishes. When $i=j$ it also vanishes, because of the point-splitting regularization implied.
The third is:
\begin{equation}\label{Weyl_3}
    \int d^2w \frac{\bpd z \pd \bar z}{|z|^2} = \int d^2w \frac{1}{|z|^2} \sum_{i,j=1}^n \frac{Q^i \bar Q^j}{(w-w_i)^2(\bar w- \bar w_j)^2}.
\end{equation}
This integral is difficult to evaluate, even when $n=2$. It has no IR-divergence. It has UV-divergences coming from the numerator and denominator. These divergences can be eliminated by renormalization of each operator, by a factor dependent at most on the momenta. The numerator gives poles at $w_i$ for all $i$. As $w$ approaches $w_i$, the integrand becomes $|w-w_i|^{-2}$, which gives a logarithmic divergence $-2\pi\ln(\ep)$ (the integral near $w_i$ of terms with $j\neq i$ in the sum vanishes due to point-splitting). The denominator gives poles at the zeros of $\bar z$, which we denote as $p_i$. The residue at these poles can be computed as follows. Let us assume that all the zeros are simple zeros, which is true for all values of $Q_i,w_i,\bar\Om,\bar\phi$, except for a zero-measure subset. Then $\bar z = (w-p_i)f(w)$, with $f(p_i)\neq 0$, $\frac{\pd \bar z}{\bar z} \rightarrow \frac{1}{w-p_i}$ as $w \rightarrow p_i$, and the residue is $1$\footnote{The sum of all residues of $\frac{\pd \bar z}{\bar z}$ vanishes (at $w \rightarrow w_i$ one has $\frac{\pd \bar z}{\bar z} \rightarrow -1$), consistent with the fact that it decays as $w^{-3}$ as $|w| \rightarrow \infty$.} Each pole will contribute another logarithmic divergence $-2\pi\ln(\ep)$ to the integral of $\left|\frac{\pd\bar z}{\bar z}\right|^2$. For the remaining finite part, \eqref{Weyl_3} can be organized as follows:
\begin{equation}\label{Weyl_4}
\begin{split}
    \int d^2w \left|\frac{\pd\bar z}{\bar z}\right|^2  = 
    & \int d^2w \left( \left|\frac{\pd\bar z}{\bar z}\right|^2 - \sn \left(\frac{1}{|w-w_i|^2} - \frac{1}{|w-w_i|^2 + r^2} + \frac{1}{|w-p_i|^2} - \frac{1}{|w-p_i|^2 + r^2} \right) \right) + \\ & \int d^2w \left(\sn \left(\frac{1}{|w-w_i|^2} - \frac{1}{|w-w_i|^2 + r^2} + \frac{1}{|w-p_i|^2} - \frac{1}{|w-p_i|^2 + r^2} \right) \right) \\
    \equiv & I_0 + I_1,
\end{split}
\end{equation}
where $r$ is some arbitrary scale. Both $I_0$ and $I_1$ have no IR-divergences. $I_0$ is finite and depends on the momenta $Q^i$ and the positions $w_i$ (as well as on $r$). $I_1$ contains the UV-divergences and is independent of the momenta $Q^i$. It can be evaluated to give $2\sn  4\pi\ln \left( \frac{r}{\ep} \right)$. In conclusion, $W$ takes the form:
\begin{equation}\label{Weyl_5}
    W = \frac{c}{48\pi} \left( \sn 8\pi\ln \left( \frac{r}{\ep} \right) + I_0 - 2\pi n\right).
\end{equation}
Upon exponentiation $e^W$ gives:
\begin{equation}\label{Weyl_6}
    e^{\frac{cI_0}{48\pi} - \frac{cn}{24}}\pn \left( \frac{r}{\ep} \right)^\frac{c}{6},
\end{equation}
where a divergent part $\ep^{-\frac{c}{6}}$ is present for each operator.

Combining \eqref{UV_4}, \eqref{UV_5} and \eqref{Weyl_6} suggests that each operator should be renormalized by:
\begin{equation}\label{UV_6}
    \OO_i(q_i) \rightarrow \hat{\OO}_i(q_i) \equiv (\mu\ep)^{-2\pi|\La||Q^i|^2} \left(\frac{\ep}{|Q^i|}\right)^{2-\Delta_i}\left( \frac{\ep}{r} \right)^\frac{c}{6} \OO_i(q_i),
\end{equation}
where $\mu$ is a renormalization scale (one can choose $r\mu = 1$ for simplicity). The first factor in \eqref{UV_6} appeared also in Cardy's work \cite{Cardy:2019qao}, see the discussion in section \ref{sec_dis} below. The second factor seems odd, since it diverges in the limit $|Q^i| \rightarrow 0$. However, the correlation function will have a well-defined limit in this case. It is worth mentioning that, similar to \cite{Cardy:2019qao}, we find a momentum-dependent renormalization of local operators, which is not possible in local QFTs.

Using \eqref{UV_6} inside \eqref{cor_dev_n_1}, with the form of the WS area in \eqref{vs_n_6}, gives the renormalized correlator:
\begin{equation}\label{cor_dev_n_2}
    \begin{split}
        Z_0 |\La| \ATS e^{|\La|\ATS} e^{-\frac{nc}{24}} \de\left(\sn q_i\right) 
        & \int\pn d^2\sg_i \de(\sg_1) e^{i\sum_{i=1}^n q_i\cdot \sg_i} C^0(\{\sg_i\})
        \prod_{i<j} \left( {\frac{|\sg_{ij}|}{2}} \right) ^{4\pi|\La|\de^{ab}Q^i_aQ^j_b} \cdot
        \\
        & \prod\limits_{i=1}^n \left(\mu^{-2\pi|\La||Q^i|^2} \right) e^{\frac{cI_0}{48\pi}} 
        \int d\bar{\Om}d\bar{\phi} D(z,0) e^{-|\La|\ATS \left( e^{2\bar{\Om}} + 2ie^{\bar{\Om}}\cos(\bar{\phi}) \right)}.
    \end{split}
\end{equation}
As in subsection \ref{subsec_reg_IR}, the limit $|\La|\ATS\to\infty$ gives a saddle point in the $\bar{\Om},\bar{\phi}$ integral. In addition, invoking momentum conservation allows combining the two factors with powers of momentum in \eqref{cor_dev_n_2}. Thus, \eqref{cor_dev_n_2} becomes: 
\begin{equation}\label{cor_dev_n_3}
    Z_0 \pi \de\left(\sn q_i\right) 
    \int \pn d^2\sg_i \de(\sg_1) e^{i\sum_{i=1}^n q_i\cdot \sg_i} C^0(\{\sg_i\})
    D(z,0) e^{\frac{cI_0}{48\pi}}
    \prod_{i<j} \left( {\frac{\mu|\sg_{ij}|}{2}} \right) ^{4\pi|\La|\de^{ab}Q^i_aQ^j_b}.
\end{equation}
Hence, the deformed correlator is the ordinary Fourier transform of the undeformed correlation function times an additional factor. This factor reflects the $\TT$-deformation, and will be dominant at large momentum. The determinant ratio $D(z,0)$ may be also regularized, but it does not seem to have UV-divergences (as can be seen in a large $|\La|$ expansion).

Returning to the two-point correlator, it can be seen that \eqref{cor_dev_9} is a special case of \eqref{cor_dev_n_3}. Ignoring the $Z_0\pi$ factor, and integrating $\sg_1$, gives the final result:
\begin{equation}\label{cor_dev_10}
        \langle \hat{\OO}(q)\hat{\OO}(-q+k)\rangle \propto
        \de(k) \int d^2\sg e^{iq\cdot \sg} e^{\frac{cI_0}{48\pi}} D(z,0)  \frac{C}{|\sg|^{2\Delta}} \left(\frac{\mu|\sg|}{2}\right)^{-\frac{q^2}{\pi|\La|}}.
\end{equation}
It should be noted that the analytic continuation of $\bar{e}$ in subsection \ref{subsec_PI_TS} causes the integral in \eqref{cor_dev_10} to have a UV-divergence as $\sg \to 0$ (for $2\Delta+\frac{q^2}{\pi|\La|}-1>1$, and in particular for large momentum). Otherwise the exponent $-\frac{q^2}{\pi|\La|}$ would have an opposite sign.

% ------------------------------------
\subsection{The Large Momentum Limit}\label{subsec_reg_hml}

% ------------
\subsubsection{Two-Point Correlator}

In the large momentum limit, one expects a saddle point approximation for the integral in \eqref{cor_dev_10}. The momentum dependencies of the different factors are:
\begin{enumerate}
    \item $\frac{C}{|\sg|^{2\Delta}}$ - Does not depend on the momentum. 
    
    \item $D(z,0)$ - As mentioned at the end of subsection \ref{subsec_PI_vs}, at large momentum $D(z,0)$ does not depend on the momentum.
    
    \item $e^{iq\cdot \sg},\left(\frac{\mu|\sg|}{2}\right)^{-\frac{q^2}{\pi|\La|}}$ - These two factors grow exponentially with the momentum.
    
    \item $e^{\frac{cI_0}{48\pi}}$ - This factor grows polynomially with the momentum.
\end{enumerate}
Therefore, one looks for the saddle of the integral:
\begin{equation}\label{hml_1}
    \int d^2\sg\ e^{iq\cdot \sg} \left(\frac{\mu|\sg|}{2}\right)^{-\frac{q^2}{\pi|\La|}} = \left(\frac{2}{\mu}\right)^{\frac{q^2}{\pi|\La|}} \int d^2\sg e^{iq\cdot \sg} |\sg|^{-\frac{q^2}{\pi|\La|}}.\footnotemark
\end{equation}
\footnotetext{Since the momentum is dimensionful, the large momentum limit is taken with respect to any dimensionful quantity (except for the UV-regulator). In the case of the plane, the only other dimensionful quantity is the deformation parameter $\La$. Hence, the large dimensionless parameter here is $N \equiv \frac{q^2}{|\La|}$. Via the change of variables $\sg^\al = \frac{|q|}{|\La|}x^\al$, the integral in \eqref{hml_1} becomes:
\begin{equation}
    \int d^2x e^{N \left(i\hat q \cdot x -\frac{1}{2\pi}\ln\left(x^2\right) \right)},
\end{equation}
where $\hat q$ is a unit vector directed along $q$. This is a suitable form for a saddle point approximation, and will be used in the following for the $n$-point correlator. It is easily verified that the saddle of this integral is the same as given below.}

\eqref{hml_1} resembles the Fourier transform of the undeformed CFT, interpreting $\sg$ as the coordinates in its flat space-time. Here the ``conformal dimension'', which started from $\Delta$ in the IR, depends solely on the momentum in the UV. Ignoring for a moment the divergence at $\sg=0$\footnote{The integration variable $\sg$ is $\sg_{21}$, and the singularity reflects the possibility for the two operator insertions to coincide. This situation is already prevented by the point-splitting regularization, as the space is cut at radius $\ep$ around each operator.}, one needs to find the saddle of $iq\cdot \sg-\frac{q^2}{\pi|\La|}\ln(|\sg|)$. This is at the complex value $\sg=-i\frac{q}{\pi|\La|}$. The determinant around the saddle is $\left(\sqrt{\frac{2\pi}{\pi|\La|}}\right)^2=\frac{2}{|\La|}$, and is independent of the momentum\footnote{in any case it does not contribute to the leading order}. It just restores the units of $d^2\sg$. Then, the value of the integral in \eqref{hml_1} is approximately (up to a phase which will be discussed in the next subsection):
\begin{equation}\label{hml_2}
    \frac{2}{|\La|} \left(\frac{|q|}{\pi e|\La|}\right)^{-\frac{q^2}{\pi|\La|}}.
\end{equation}
The result in \eqref{hml_2} should be accompanied by substituting the saddle in the rest of the integrand. The factor of the undeformed correlator in \eqref{cor_dev_10} gives:
\begin{equation}\label{hml_3}
    \frac{1}{|\sg|^{2\Delta}} \rightarrow \left(\frac{\pi|\La|}{|q|}\right)^{2\Delta},
\end{equation}
which gives a factor of $|q|^{-2\Delta}$. Putting together \eqref{hml_2} and \eqref{hml_3} yields:
\begin{equation}\label{hml_4}
    \frac{2}{|\La|} \left(\frac{|q|}{\pi e|\La|}\right)^{-\frac{q^2}{\pi|\La|}}
    \left(\frac{\pi|\La|}{|q|}\right)^{2\Delta} = \frac{1}{|\La|}\frac{2}{e^{2\Delta}} \left(\frac{\pi e|\La|}{|q|}\right)^{\frac{q^2}{\pi|\La|}+2\Delta}.
\end{equation}
Including the factor from \eqref{hml_1} and the $C$ from the undeformed correlator gives us the final result: 
\begin{equation}\label{hml_5}
    \frac{2C}{|\La|} \left(\frac{\mu}{2e}\right)^{2\Delta} \left(\frac{2\pi e|\La|}{\mu|q|}\right)^{\frac{q^2}{\pi|\La|}+2\Delta},
\end{equation}
where we gathered all momentum-dependent terms. 

It is natural to be suspicious of this result coming from an imaginary saddle point, both because it ignores the divergence at $\sg=0$ (which can be regulated by point-splitting as was done elsewhere), and because it is exponentially small at large momentum, so it is not clear that it gives the dominant contribution. In appendix \ref{sec_fur} a more careful analysis of the Fourier transform is performed, and a very similar result to \eqref{hml_5} is found. The only difference is that the factor of $2/e^{2\Delta}$ above is replaced by an oscillating momentum-dependent factor $1/\sin\left(\frac{q^2}{2|\Lambda|} + \pi\Delta\right)$ (which is subleading at large momentum), such that the two-point correlator is given by
\begin{equation}\label{hml_6}
    \frac{C}{|\La|}\frac{1} {\sin\left(\frac{q^2}{2|\Lambda|} + \pi\Delta\right)} \left(\frac{\mu}{2e}\right)^{2\Delta} \left(\frac{2\pi e|\La|}{\mu|q|}\right)^{\frac{q^2}{\pi|\La|}+2\Delta}.
\end{equation}

It should be noted that (up to the fluctuating coefficient) the leading result in the UV limit does not depend on $\Delta$, i.e. the initial information of the undeformed CFT in the IR is lost. As will be shown momentarily, this property is also true for the $n$-point function.

% ------------
\subsubsection{n-Point Correlator}

For discussing the large momentum limit of the $n$-point correlator, we define: 
\begin{equation}\label{hml_n_1}
    q_i \equiv \sqrt{\pi}{K} y_i,  
\end{equation}
where $K$ is a typical momentum scale in the correlator, $N \equiv K^2 / |\La| >> 1$, and $|y_i|$'s are of order one. The $\sqrt{\pi}$ is introduced for later convenience. We do not impose a hierarchy inside the momenta themselves, but we do assume that $N$ is very large with respect to any other dimensionless quantity, in particular the ratio of two $y_i$'s or any polynomial of them.

All of the considerations of the last subsection about the terms contributing to the saddle for the two-point correlator are also valid for the expression in \eqref{cor_dev_n_3} for the $n$-point correlator\footnote{In particular, the determinant ratio $D(z,0)$, involving $\Om$ given in \eqref{vs_n_3}, does not depend on $N$ in the large $N$ limit.}. Therefore, we seek the saddle of the following integral:
\begin{equation}\label{hml_n_2}
    \int \pn d^2\sg_i \de(\sg_1) e^{i\sum_{i=1}^n q_i\cdot \sg_i} \prod_{i<j} \left( {\frac{\mu|\sg_{ij}|}{2}} \right) ^{\frac{q_i\cdot q_j}{\pi|\La|}}.
\end{equation}
The saddle of $\{\sg_i\}$ is on the imaginary axis, and the change of variables $\sg_i \equiv i\sqrt{\frac{N}{\pi|\La|}}x_i$ gives \eqref{hml_n_2} a form suitable for a saddle point approximation: 
\begin{equation}\label{hml_n_3}
\begin{split}
    & \int \pn d^2\sg_i \de(\sg_1) e^{i\sum_{i=1}^n q_i\cdot \sg_i +\sum_{i<j} {\frac{q_i\cdot q_j}{2\pi|\La|}}\ln\left( \left(\frac{\mu\sg_{ij}}{2}\right)^2 \right)} \propto \\
    & \int \pn d^2x_i \de(x_1) e^{N \left( -\sum_{i=1}^n y_i\cdot x_i +\sum_{i<j} \frac{y_i\cdot y_j}{2} \ln \left(-\frac{N}{\pi|\La|}\left(\frac{\mu x_{ij}}{2} \right)^2 \right) \right)}.
\end{split}
\end{equation}
($x_i,y_i$ are two-dimensional vectors.) Differentiating the exponent with respect to $x_i$ gives $n$ equations for the saddle point:
\begin{equation}
    \sum_{j\neq i}(y_i\cdot y_j) \frac{x_{ij}}{x_{ij}^2} = y_i.
\end{equation}
These equations can not be solved easily. However, for the saddle, only two properties of the solution are important: that it exists, and that it scales linearly with $y_i$ (i.e. if $y_i \to a y_i$ then $x_i \to a x_i$). Then, as long as $|y_i|<<N$ also $|x_{ij}|<<N$. Therefore, the (logarithmic) dominant part of the saddle is:
\begin{equation}\label{hml_n_4}
    N^{N\frac{1}{2} \sum_{i < j}y_i\cdot y_j},
\end{equation}
which does not depend on the saddle point value of $x_i$ at all. Momentum conservation $\left( \sum_{i=1}^n y_i \right)^2 = 0$ allows to write the final result as:
\begin{equation}\label{hml_n_5}
    \left( \frac{{K}^2}{|\La|} \right)^{-\sn \frac{q_i^2}{4\pi |\La|}},
\end{equation}
generalizing the two-point correlator \eqref{hml_2}. As in our discussion of the two-point correlator above, even though the saddle point gives an exponentially small result, the exact result is expected to be close to this\footnote{Up to terms similar to the $\csc$ factor in the two-point correlator \eqref{hml_6}.}.

It remains to show that $I_0$ depends only logarithmically on $N$, and therefore, it will not change the saddle point found. The integrand $|\pd\ln(\bar z)|^2$ has $2n$ poles (see \eqref{Weyl_4} and the discussion around it in subsection \ref{subsec_reg_UV}), and the integral $I = \int d^2w|\pd\ln(\bar z)|^2 = I_0 + I_1$ became finite by a point-splitting regularization, i.e. by cutting a disk of radius $\ep$ around each pole. This gives a function $I({w_i},e^{-\bar\Om}{Q^i},\ep)$ \footnote{$\left| e^{\bar\Om} \right| = |i| $ is of order unity as found after \eqref{IR_1}}, which satisfies:
\begin{equation}
    I\left(\{xw_i\},e^{-\bar\Om}\{Q^i\},\ep\right) = I\left(\{w_i\},e^{-\bar\Om}\{\frac{Q^i}{x}\},\frac{\ep}{x}\right).
\end{equation}
For $Q^i =\frac{1}{2}\sqrt{\frac{N}{|\La|}}y^i$, and $w_i = \sqrt{\frac{N}{\pi|\La|}}z_i$, one has: 
\begin{equation}
    I\left(\sqrt{\frac{N}{\pi|\La|}}\{z_i\},\frac{e^{-\bar\Om}}{2}\sqrt{\frac{N}{|\La|}}\{y_i\},\ep\right) = I\left(\sqrt{\frac{1}{\pi}}\{z_i\},\frac{e^{-\bar\Om}}{2}\{y_i\},\sqrt{\frac{|\La|}{N}}\ep\right).
\end{equation}
The form of the divergences dictates the dependence on $\ep$ to be $-\sum_{i=1}^{n} 8\pi\ln(\ep) + O(1)$, hence this rescaling yields a term of $-\sum_{i=1}^{n} 8\pi\ln\left(\sqrt{\frac{|\La|}{N}}\ep\right)$. For $I_0 = I - I_1$, where $I_1 = \sum_{i=1}^{n} 8\pi\ln\left(\frac{r}{\ep}\right) + O(\ep)$, in the limit $\ep\to 0$, one finds the dependence $I_0 = f\left(\{z_i\},e^{-\bar\Om}\{y_i\}\right) - \sum_{i=1}^{2n} 4\pi\ln\left(\sqrt{\frac{|\La|}{N}}r\right)$. This $\ln(N)$ dependence is subleading to the $N$ dependence found in the saddle, and it will not affect the large $N$ limit.

Last point to mention regards the minus sign inside the logarithm in \eqref{hml_n_3}. Using momentum conservation, it gives an overall phase of:
\begin{equation}
    (-1)^{-\sum_{i=1}^n \frac{q_i^2}{4\pi |\La|}} = e^{-i\pi\sum_{i=1}^n \frac{q_i^2}{4\pi |\La|}} = \pn e^{-\frac{i q_i^2}{4|\La|}}.
\end{equation}
There may be other contributions to this phase, but this is the dominant contribution in the large momentum limit.
The appearance of this phase is strange, since in general, the correlation function in momentum space is expected to satisfy $C(q_1,q_2)^* = C(-q_1,-q_2)$. However, the phase factorizes, so it can be swallowed in the definition of the renormalized operator $\hat{\OO}$.

% ====================================
\section{Comparison to Previous Results}\label{sec_dis}

Our final result \eqref{hml_6} for the two-point correlator in the large-momentum limit is quite different from Cardy's result \eqref{cor_Crdy} mentioned in the introduction, and in this section we compare the two in detail. For the comparison, the same notations should be used. In Cardy's paper \cite{Cardy:2019qao}, the regulator $\La$ in his notations (which will be denoted by $\La_c$ to distinguish it from our deformation parameter $\La$) is the same as what we called $\ep^{-1}$ here, and the deformation parameter $2\pi \lambda$ in his notation (which will be denoted by $2\pi \lambda_c$) is $t$ here. Using \eqref{INT_5}, one gets the relation $2\pi\lambda_c=\frac{1}{|\La|}$. 

First, we compare the renormalization of the operators in momentum space. It is given in (1.6) in \cite{Cardy:2019qao}, where the undeformed fields were multiplied by the factor:
\begin{equation}\label{dc_1}
    e^{\lambda_c \ln\left(\frac{\La_c}{\mu}\right)q^2} = \left(\frac{\La_c}{\mu}\right)^{\lambda_c q^2} \rightarrow (\ep\mu)^{-\frac{q^2}{2\pi|\La|}}
\end{equation}
(where the arrow denotes translation to our notations). This is exactly the first factor in our renormalization in \eqref{UV_6}, remembering that $|Q|=\frac{|q|}{2\pi|\La|}$. The second and third factors in \eqref{UV_6} are present only in the JT-formalism. This is reasonable, since these factors were present because of an auxiliary WS and a metric on it. The second factor was needed to take care of the divergence caused by the factor $\sqrt{g(\sg_i)}$ in the operators' definition in \eqref{op_def_mom}. This factor made our operators diff-invariant. The third factor was needed because of the Weyl anomaly, present because the WS metric is not flat. In the perturbative $\TT$-formalism, there is no equivalent notion of the WS which appears in the JT-formalism. There, the (bare) operators in momentum space can be simply defined as the Fourier transform of the operators $O(\sg)$. In general, the relation between the operators we define and the operators of the perturbative $\TT$-formalism is not clear (though they agree to leading order in the deformation). In appendix \ref{sec_diff_op} we suggest an alternative definition of the operators in our formalism (with the same large-momentum behavior for their correlators), which in a specific gauge (the gauge $X^a = \sg^a$) looks identical to the perturbative definition, but these operators still live on a space with a curved metric, so it is not clear how to compare them to the original formulation.

Next, we compare the large momentum limit of the renormalized correlation functions. It is given in (4.27) in \cite{Cardy:2019qao} as (denoting $\Delta$ in his notation as $\Delta_c$):
\begin{equation}\label{dc_2}
     |q|^{2\Delta_c}\left(\frac{|q|}{\mu}\right)^{2\lambda_c q^2} \rightarrow |q|^{2\Delta-2}\left(\frac{|q|}{\mu}\right)^{\frac{q^2}{\pi|\La|}}
\end{equation}
Cardy used $\Delta_c$ as the momentum-space conformal dimension of the undeformed theory, and we use $\Delta$ as the position-space conformal dimension of the undeformed theory, so $2\Delta_c = 2\Delta-2$. Four differences between \eqref{dc_2} and our result in \eqref{hml_6} should be noted:
\begin{enumerate}
    \item The most important is in the sign of the exponent, before the common factor of $\frac{q^2}{\pi|\La|}$. While Cardy's result diverges as the momentum goes to infinity, the result we found tends to zero. This difference appears despite the same sign appearing in the renormalization of the operators $(\ep\mu)^{-\frac{q^2}{2\pi|\La|}}$. The two-point correlator in position space is finite at small distances, due to the fast decay in momentum space. 
    
    \item The dimensionful quantity $q^{-2}$ in Cardy's formalism made the appearance of $|q|^{2\Delta}$ become the undeformed correlator $C^0(q)$. In our result, the dimensions were taken care of  by a factor of $|\La|^{-1}$. In the large momentum limit this difference is small and our approximations are not sensitive to it.
    
    \item The basis of the exponent. While in \eqref{dc_2} the momentum $q$ is multiplied by the inverse of the renormalization scale $\mu$ and becomes dimensionless, in \eqref{hml_6} it is multiplied by $\frac{1}{2\pi e}\frac{\mu}{|\La|}$. This difference can be eliminated by using a different renormalization scale $\mu \rightarrow \frac{2\pi e|\La|}{\mu}$, although doing so will make a difference in the renormalization factor of the operators in \eqref{dc_1}. In any case, we are again not sensitive to this difference in the large-momentum approximation made.
    
    \item An extra factor of 
    $1/\sin\left(\frac{q^2}{2|\Lambda|} + \pi \Delta\right)$ (from the Fourier transform).
\end{enumerate}

The main reason for these differences is that Cardy's result was based on computing the leading logs in the correlation function at small momentum and resumming them. We find precisely the same leading logs at small momentum in our result, as shown for the first order expansion in $\frac{q^2}{|\La|}$ in appendix \ref{sec_hgh_lam}. This can be shown to hold more generally for the leading logs at each order. The determinant ratio $D(z,0)$ in \eqref{cor_dev_10} does not contain logs, and therefore, all the dominant logs come from the Fourier transform:
\begin{equation}
    \langle \hat{\OO}(q)\hat{\OO}(-q+k)\rangle \propto
    \de(k) \int d^2\sg e^{iq\cdot \sg} \frac{C}{|\sg|^{2\Delta}} \left(\frac{\mu|\sg|}{2}\right)^{-\frac{q^2}{\pi|\La|}}.
\end{equation}
Using the general result for the Fourier transform of a power law developed in appendix \ref{sec_fur}, one gets the result (see \eqref{fur_3}):
\begin{equation}
    \frac{1}{\Gamma(\bb)^2}\left(\frac{|q|}{2}\right)^{2\bb-2} \ , \ \bb \equiv \frac{q^2}{2\pi|\La|} + \Delta.
\end{equation}
This result is essentially \eqref{dc_2}, up to the $\Gamma(\bb)^{-2}$ factor. This extra factor, when expanded in small momentum, does not contribute logs, and therefore, the leading logs of our correlator will be the same as the leading logs of \eqref{dc_2}. However, we claim that the large momentum result is different, and is not dominated by the leading logs. This is because the convergence radius of $\Gamma(\bb)^{-2}$ is finite, due to poles at negative values of $q^2$. In the large momentum limit, this term dominates, and flips the sign of the exponent to give the result in \eqref{hml_6}, as shown in appendix \ref{sec_fur}. This is summarized in figure \ref{fig_cor}. Therefore, it seems that using our definition of the operators the result in \cite{Cardy:2019qao}, found by re-summation of leading logs, is valid only for small momentum.
\begin{figure}
\captionsetup{singlelinecheck = false, justification=justified}
\includegraphics[width=\textwidth]{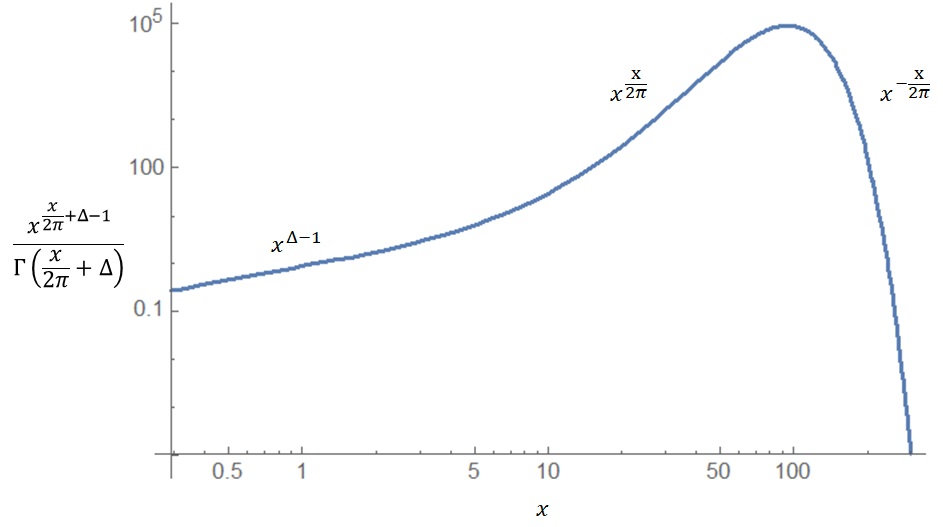}
\caption{Schematic dependence of the logarithm of the two-point correlator on the logarithm of the momentum, showing three different regions. $x \equiv q^2/|\La|$, and $\Delta = 2$. The first region is the very low-momentum result, dictated by the undeformed CFT correlator $q^{2\Delta-2}$. The second is the summation of the leading logs, leading to a growing function. The third is the large momentum limit.}
\label{fig_cor}
\end{figure}

Note that we could swallow the leading large momentum dependence into the operator normalization by multiplying each operator by $|q_i|^{+\frac{q_i^2}{2\pi|\La|}}$ (in addition to the factors appearing in \eqref{UV_6}). This would cancel the leading terms in all $n$-point function we compute, as can be seen using \eqref{hml_n_5}:
\begin{equation}
    \frac{\left\langle \pn\hat{\OO}_i(q_i) \right\rangle}{\pn\sqrt{\left\langle \hat{\OO}_i(q_i)\hat{\OO}_i(-q_i) \right\rangle}} \xrightarrow[{N \rightarrow \infty}]{} \frac{N^{-N\sum_{i=1}^n \frac{y_i^2}{4}}}{\pn\sqrt{\sqrt{N}^{-N y_i^2}}} ,
\end{equation}
where $q_i = \sqrt{\pi|\La|N}y_i$ and $|y_i|$ are of order unity. One can see that the numerator cancels the denominator. Since the renormalization of the operators in these theories is momentum-dependent, there is an ambiguity in multiplying each operator by a function of the momentum, and it is not clear how to fix this ambiguity in a non-local theory. In this paper we chose a minimal renormalization prescription where the operators are renormalized just by the factors needed to remove the UV-divergences encountered. As we mention below, the results in this normalization agree with results from string theory where the normalization is determined by completely different methods. It would be interesting to understand if there is any preferred choice of the normalization of the operators in these non-local theories. Of course, appropriate ratios of correlation functions do not depend on the normalization and are free from this ambiguity.

We note that in \cite{Asrat:2017tzd} there is a somewhat similar calculation of a two-point correlation function under an irrelevant deformation which resembles the $\TT$-deformation in some ways (it is the ``single-trace'' version of the $\TT$-deformation, when applied to symmetric orbifold CFTs). There, in (3.10), the correlator at large momentum behaves as:
\begin{equation}\label{dc_3}
    \sim \left( \left(a\lambda_k\mu^2\right)^{\sqrt{a}} \frac{\lambda_k q^2}{a} \right) ^ {-\sqrt{\frac{\lambda_k q^2}{a}}},
\end{equation}
where $\lambda_k$ is the deformation parameter, $\mu$ is a renormalization scale, and $a$ is some number. There is a little similarity to our result in \eqref{hml_6}, in the sense that in both cases the correlation function decays at large momentum, unlike Cardy's result in \eqref{dc_2}. But in \eqref{dc_3} the correlator decays with $|q|$ in the exponent, unlike $q^2$ in our case. In any case, the physical theory, its deformation, and the definition of the renormalized operators involved in \cite{Asrat:2017tzd} are quite different than the ones considered here, so it is not clear if the two results should be similar.

Just before this paper appeared, another string theory correlation function in the string background corresponding to the ``single-trace'' $\TT$-deformation was computed in \cite{Cui:2023jrb}. In that paper, the two-point function was computed for so-called ``long string'' states that wind around the AdS geometry that appears in the IR limit of the corresponding background. These states are known to be well-described (away from the far-IR region of the geometry) by a symmetric product orbifold $M^{Q_1}/S_{Q_1}$ in the CFT dual to the original $AdS_3$ background (see \cite{Chakraborty:2019mdf} and references therein), and performing the ``single-trace'' $\TT$-deformation on the original CFT corresponds to performing the standard ``double-trace'' $\TT$-deformation on the factor $M$ appearing in this symmetric product. For ``long string'' states with ``winding number'' $w=1$ in this background, corresponding to the untwisted sector of the orbifold, one can then argue that their correlation functions should be the same as those in the CFT $M$ deformed by the standard $\TT$-deformation (at least for large momenta). The two-point correlation function of these states was computed in \cite{Cui:2023jrb}, and the result they found is exactly the same as the result \eqref{hml_6n} that we found for the two-point function using the JT-gravity formulation of these theories. This gives us a consistency check on both computations, and it suggests that the definition of the operators in the $\TT$-deformed CFTs using their string theory holographic dual agrees with the definition in JT-gravity that we used above (at least for large momentum). It would be interesting to study further the relation between these two very different approaches, and in particular to see if the high-energy behavior of other correlation functions also agrees.

Another approach to defining operators and correlation functions for J$\bar{\textrm{T}}$-deformed theories was suggested in \cite{Guica:2021fkv}, and can presumably be generalized also to $\TT$-deformed theories. Another possibility for defining operators in the $\TT$-deformed theories was given in \cite{Kruthoff:2020hsi}. It would be interesting to understand the relation between these operators and the ones used here. Note that using the formalism of \cite{Kruthoff:2020hsi} on the plane, the $\TT$ deformation can be viewed as a canonical transformation, and in that language one can also define deformed operators by a canonical transformation, such that they would have the same correlation functions as in the undeformed theory. However, that definition would not agree with our definition even at leading order in the deformation, and presumably gives very complicated operators in our language.

There are various interesting generalizations of our work and future directions. It will be interesting to consider correlation functions on the torus instead of the plane, where there are no IR-divergences; naively we expect that the large momentum limit of the two-point correlator will be similar to \eqref{hml_6}, but in a non-local theory this is far from obvious. It would be interesting to understand more precisely the large-momentum behavior of $n$-point correlation functions, and how this affects the short-distance behavior of the $\TT$-deformed theories. It was noted in \cite{Dubovsky:2017cnj} that the S-matrix in the decoupled theories on large $N$ QCD strings is similar to that of $\TT$-deformed theories, and it would be interesting to check if this similarity extends also to the correlation functions computed here.

It would also be interesting\footnote{We thank S. Dubovsky, G. Hernandez-Chifflet and Z. Komargodski for this suggestion.} to understand if there is any LSZ-type relation between the correlator computed in this paper and the (known) S-matrix of the $\TT$-deformed theories, and to compute the correlator in the relevant Lorentzian regime for this (which is not directly related to the Euclidean regime we work in, so it is not clear how this computation should be performed).

% ====================================
\section*{Acknowledgments}

We would like to thank Sergei Dubovsky, Guzmán Hernández-Chifflet, Zohar Komargodski, David Kutasov, Onkar Parrikar, and Adar Sharon for useful discussions. This work was supported in part by an Israel Science Foundation (ISF) center for excellence grant (grant number 2289/18), by ISF grant no. 2159/22, by Simons Foundation grant 994296 (Simons
Collaboration on Confinement and QCD Strings), by grant no. 2018068 from the United States-Israel Binational Science Foundation (BSF), by the Minerva foundation with funding from the Federal German Ministry for Education and Research, by the German Research Foundation through a German-Israeli Project Cooperation (DIP) grant ``Holography and the Swampland'', and by a research grant from Martin Eisenstein. OA is the Samuel Sebba Professorial Chair of Pure and Applied Physics.

\appendix

% ====================================
\section{The Faddeev-Popov Determinant}\label{sec_FP}

This appendix is devoted to the FP determinant. We write the naive gauge-fixing (subsection \ref{subsec_FP_gf}), then explicitly calculate the corresponding determinant (subsection \ref{subsec_FP_gft1}). The determinant vanishes, reflecting the fact that the naive gauge-fixing did not completely fix the gauge (subsection \ref{subsec_FP_zero}). Then, the gauge will be completely fixed, presenting the gauge-fixing used in \eqref{FP_id_3}, and the determinant used in \eqref{det_dec_1} will be calculated (subsection \ref{subsec_FP_gft2}).

% ------------------------------------
\subsection{Gauge Fixing}\label{subsec_FP_gf}

The diff-invariance present in the PI calculating the correlation function will be gauge-fixed by the usual FP procedure. For this one needs to impose constraints on some coordinate-dependent quantity. In principle, the $X^a(\sg)$ can be chosen to be some simple function of the coordinates, say $X^a(\sg) = I^a_{\al}\sg^{\al}$ (``static gauge''). However, the $X^a$ PI is linear and thus easy to perform, and so instead the vielbein will be gauge-fixed with the orthogonal.
By using the two diffeomorphisms, any metric on the two-dimensional infinite plane can be brought to the form:
\begin{equation}\label{fx_mtrc_sg_pl}
    [\bar g_{\al\bb}]_{\sg}=e^{2\Om(\sg)}
    \de_{\al\bb}.
\end{equation}
The vielbein then takes the form:  
\begin{equation}        
    [\bar{e}^a_{\al}]_{\sg} = e^{\Om(\sg)}e^{\ep\phi(\sg)}I^a_\al.
\end{equation}
The FP identity is:
\begin{equation}\label{FP_id_1}
    1 = \int\DD V\DD\Om\DD\phi J(e) \de\left(e^{(V)} - \bar e(\Om,\phi)\right).
\end{equation}

% ------------------------------------
\subsection{Determinant Evaluation - I}\label{subsec_FP_gft1}

The determinant $J(\bar{e})$ need not be a constant, and can depend on $\Om,\ \phi$\footnote{One can guess that it will depend only on $\Om$. This is because the determinant relates volumes in the infinite dimensional manifold with coordinates $V^{\al}(\sg),\Om(\sg),\phi(\sg)$, and the metric on this manifold depends only on $\Om$.\label{foot_det_dep}}. Considering the constraint:
\begin{equation}
    {\hat C}(V^{\al}(\sg),\Om(\sg),\phi(\sg),e^a_{\al}(\sg)) \equiv e^a_{\al}(\sg)^{(V^{\al}(\sg))}-\bar{e}^a_{\al}(\Om(\sg),\phi(\sg)),    
\end{equation}
we take its linearized form around the solution $\bar{e}$:
\begin{equation}\label{P_def_one}
    \begin{split}
        P&\equiv\frac{\de {\hat C}}{\de (V ,\Om,\phi)},\\
        P\begin{pmatrix}
        \de V^{\bb} \\ \de\Om \\ \de\phi
        \end{pmatrix}
        &=
        \begin{pmatrix}
            \pd_{\bb}\bar{e}^a_{\al} + \bar{e}^a_{\bb}\pd_{\al}& 
            -\bar{e}^a_{\al}&
            -\ep^a_b\bar{e}^b_{\al}
        \end{pmatrix}
        \begin{pmatrix}
        \de V^{\bb} \\ \de\Om \\ \de\phi
        \end{pmatrix},
        \\
        \pd_{\bb}\bar{e}^a_{\al} &= \bar{e}^a_{\al}\pd_{\bb}\Om+\ep^a_b\bar{e}^b_{\al}\pd_{\bb}\phi.
    \end{split}
\end{equation}
The operator $P$ connects two different linear spaces: $U = \{\left(\de V^{\bb},\de\Om,\de\phi\right)\}\rightarrow E = \{\de e^a_{\al}\}$. The first, $U$, is made of three different sectors orthogonal to one another, and the second, $E$, is the tangent space to the vielbein space. 
The absolute value of the determinant of $P$, evaluated at $\bar{e}$, is $J(\bar e)$. Calculating such a determinant is usually a difficult task. Moreover, the ``dimension'' of the domain and of the range of $P$ need not be the same, so this determinant is in fact not well-defined\footnote{Although naive counting will give the same number of dimensions of both spaces. At each space one has four functions, defined over the coordinate space, giving $4\cdot \aleph^2$ degrees of freedom. After applying some regularization for each space the dimensions can be compared more precisely. One has to be careful, as different regularizations may be used for $U$ and for $E$.}. To overcome the difficulties mentioned above, the operator $P^\dagger P$ is considered instead of $P$. Its advantage is that its determinant is always well-defined, as it is a function from $U$ to itself. Its determinant will be used instead of $\det(P)$, following the fact that when the latter exists it is equal to $\sqrt{\det(\Pd P)}$. 
In order to write $\Pd$, one needs to define inner products on $U, E$. We have $G_s$ for the subspaces of $\de\Om, \de\phi$, which are scalars of the WS, $G_v$ for $\de V^{\bb}$ which is a vector of the WS, and $G_e$ for $\de e^a_\al$ which are co-vectors of the WS:
\begin{equation}\label{in_prod_sve}
    \begin{split}
        G_s(\de f,\de g) & \equiv |\La|\int dA\de f(\sg) \de g(\sg),  \\
        G_v\pr{\de V^{\al},\de W^{\bb}} & \equiv |\La|\int dAg_{\al\bb}\de V^{\al}(\sg) \de W^{\bb}(\sg), \\
        G_e\pr{\de e^a_{\al},\de f^b_{\bb}} & \equiv |\La|\int dAg^{\al\bb}\de _{ab}\de e^a_{\al}(\sg) \de f^b_{\bb}(\sg).
    \end{split}
\end{equation}
$G_s,G_v,G_e$ are the simplest inner products which are diff-invariant. A factor of $|\La|$ is introduced in order to make the inner products dimensionless. $G_e$ is essentially the same as $G_v$, but with the inverse of $g_{\al\bb}$, which is suitable to contract the two lower indices of the two (infinitesimal) vielbein. The TS indices of the vielbein are treated as two independent vectors contracted with $\de_{ab}$. $G_s$ for $\de\Om, \de\phi$ results from $G_e$ for the vielbein in the orthogonal. Thus, we can write the inner products for the two spaces $U,E$:
\begin{equation}\label{in_prod_UE}
    \begin{split}
         G_E(\de e^a_{\al},\de e'^b_{\bb}) &\equiv G_e(\de e^a_{\al},\de e'^b_{\bb}), \\
         G_U\left(\left(\de V^{\al},\de\Om,\de\phi\right),(\de V'^{\bb},\de\Om',\de\phi')\right) & \equiv 
        G_v(\de V^{\al},\de V'^{\bb})+G_s(\de\Om,\de\Om')+G_s(\de\phi,\de\phi').
    \end{split}
\end{equation}
Using these inner products, one can write $\Pd : E\rightarrow U$: 
\begin{equation}\label{Pd_def}
    \Pd\de e_{a\g}=
    \begin{pmatrix}
        \bar{g}^{\bb\de}\bar{g}^{\al\g}(-\bar{e}^a_{\bb}\pd_{\al}-\pd_{\al}\bar{e}^a_{\bb}+\pd_{\bb}\bar{e}^a_{\al})\\
        -\bar{g}^{\al\g}\bar{e}^a_{\al} \\
        -\bar{g}^{\al\g}\ep^a_b \bar{e}^b_{\al}
    \end{pmatrix}
    \de e_{a\g},
\end{equation}
and calculate $\Pd P: U \rightarrow U$ by composing \eqref{Pd_def} with \eqref{P_def_one}:
\begin{equation}\label{PdP_def}
\Pd P =
    \begin{pmatrix}
        {P_{11}}^{\de}_{\bb}
            & \pd^{\de}
            & \ee^{\de\al}\pd_{\al} + 2\ee^{\de\al}\pd_{\al}\Om -2\pd^{\de}\phi \\
            -2\pd_{\bb}\Om-\pd_{\bb}
            & 2 
            & 0  \\
            -2\pd_{\bb}\phi-{\ee_{\bb}}^\al\pd_{\al} 
            & 0
            & 2
    \end{pmatrix},
\end{equation}
where
\begin{equation}
    \begin{split}
        {P_{11}}^{\de}_{\bb} \equiv & -\de^{\de}_{\bb} g^{\al\g}\pd_{\al}\pd_{\g} -(\de^{\de}_{\al}\pd_{\bb}\Om+2\de^{\de}_{\bb}\pd_{\al}\Om+\ee^{\de}_{\al}\pd_{\bb}\phi-\ee_{\bb\al}\pd^{\de}\phi)\pd^{\al}+\pd^{\de}\Om\pd_{\bb}+2\pd^{\de}\phi\pd_{\bb}\phi \\
        & -2\ee^{\de}_{\al}\pd^{\al}\Om\pd_{\bb}\phi-\pd^{\de}\pd_{\bb}\Om-\ee^{\de}_{\al}\pd^{\al}\pd_{\bb}\phi.
    \end{split}
\end{equation}
Calculating the determinant of $\Pd P$ directly is difficult, since every ``element'' is a matrix. Thus, it is easier to decompose $\Pd P$ into an LDU-type decomposition, where the determinant of each component is easy to calculate. The following decomposition is found: 
\begin{equation}\label{dec_PdP}
\Pd P=T^{\dagger}NT,
\end{equation}
with
\begin{equation}\label{dec_PdP_1}
    T=
    \begin{pmatrix}
        1
        & 0
        & 0 \\
        -\pd_{\bb}\Om-\frac{1}{2}\pd_{\bb}
        & 1 
        & 0 \\
        -\pd_{\bb}\phi-\frac{1}{2}{\ee_{\bb}}^\al\pd_{\al}
        & 0
        & 1
    \end{pmatrix} 
    ,\ 
    N=
    \begin{pmatrix}
        c^{\de}_{\bb} & 0 & 0 \\
        0             & 2 & 0 \\
        0             & 0 & 2
    \end{pmatrix}
    ,\ c^{\de}_{\bb}=-\frac{1}{2}{\de}^{\de}_{\bb}\pd^2+\pd^{\de}\Om\pd_{\bb}-\pd_{\bb}\Om\pd^{\de}-{\de}^{\de}_{\bb}\pd_{\al}\Om\pd^{\al}.
\end{equation}
Since $\Pd P$ is a Hermitian matrix, so should be $N$ and $c^{\de}_{\bb}$, as one can easily verify. With this decomposition one gets:
\begin{equation}
    J(\bar e) = \sqrt{\det(\Pd P)} = \sqrt{\det(N)} \propto \sqrt{\det(c)}
\end{equation} 
(up to $\det(2)^2$ which can be ignored). Hence, we focus on the operator $c^{\de}_{\bb}$.

There are two things worth noting.
First, $J(\bar e)$ depends only on $\Om$ (see footnote \ref{foot_det_dep}). Second, the FP determinant is diff-invariant as expected. The two-derivative terms in $c^{\de}_{\bb}$ suggest the (vectorial) Laplacian as a candidate for the diff-invariant operator, $g^{\al\g}\nabla_{\al}\nabla_{\g}\de V^{\bb}$. Evaluating it for the orthogonal gauged metric gives $c^{\de}_{\bb}$, up to one term, $g^{\al\g}\pd_{\al}\pd_{\g}\Om\de^{\de}_{\bb}$, which exists in the vectorial Laplacian and does not exist in $c^{\de}_{\bb}$. This one term can be written as a diff-invariant quantity, $-R/2$, where $R$ is the Ricci scalar evaluated in the orthogonal. Thus, $c=-\frac{1}{2}\nabla^2-\frac{1}{4}R$, and it is manifestly diff-invariant.

% ------------------------------------
\subsection{Zero Modes}\label{subsec_FP_zero}

One can see that $P$ is not invertible, because it has zero modes. Their appearance means that for a given vielbein the $\de$-function chooses more than one value out of the space of $V,\Om,\phi$, i.e. there are more than one conformal form for any initial vielbein. In fact, the whole conformal algebra is expected to be its zero modes. This can be seen by using complex coordinates, where $c^\de_\bb$ is diagonal and has the form:
\begin{equation}\label{c_com_w}
    [c]_w
    \begin{pmatrix} \de V \\ \widebar{\de V}    \end{pmatrix}
    = e^{-2\Om}
    \begin{pmatrix}
        -\pd\bpd-4\pd\Om\bpd & 0 \\ 0 & -\pd\bpd-4\bpd\Om\pd
    \end{pmatrix}
    \begin{pmatrix} \de V \\ \widebar{\de V}    \end{pmatrix}.
\end{equation}
Any holomorphic diffeomorphism $(\de V(w),0)^T$ and anti-holomorphic diffeomorphism $(0, \widebar{\de V}(\bar{w}))^T$ are thus zero modes of $c^\de_\bb$. As explained in appendix \ref{sec_boun}, the allowed boundary conditions for the diff $\de V^\bb$ exclude all conformal transformations but translations.

% ------------------------------------
\subsection{Determinant Evaluation - II}\label{subsec_FP_gft2}

We fix the remaining diffeomorphism invariance position of one of the operators to $\sg_1=0$, and thus modify the $\de$-function constraint \eqref{FP_id_1}:
\begin{equation}
    1 = \int\DD V\DD\Om\DD\phi J(e,\sg_1) \de\left(e^{(V)} -\bar e \right)\de\pr{\sg_1^{(V)}-0}.
\end{equation}
To calculate the Jacobian $J(\bar e,0)$, we extend the range of $P$ to $E\equiv\left\{(\de e^a_{\al},\de\sg_1^{\al})\right\}$. Then, \eqref{P_def_one} becomes:
\begin{equation}\label{P_def_1}
    P\begin{pmatrix}
        \de V^{\bb}(\sg) \\ \de\Om(\sg) \\ \de\phi(\sg)
        \end{pmatrix}
        =
        \begin{pmatrix}
            \pd_{\bb}\bar{e}^a_{\al}(\sg)+\bar{e}^a_{\bb}(\sg)\pd_{\al} 
            & -\bar{e}^a_{\al}(\sg) 
            & -\ep^a_b\bar{e}^b_{\al}(\sg) \\
            \de^\al_\bb \int d^{2}\sg\de(\sg) 
            & 0 
            & 0
        \end{pmatrix}
        \begin{pmatrix}
        \de V^{\bb}(\sg)\\ \de\Om(\sg) \\ \de\phi(\sg)
        \end{pmatrix}.
\end{equation}
As before, in order to define $P^\dagger$, the inner product $G_E$ must be modified to include the $\de\sg_1^\al$ sector. The natural diff-invariant choice for the vector $\de\sg_1^\al$ is
\begin{equation}\label{in_prod_S}
    G_{\sg}(\de\sg_1^{\al},\de{\sg'}_1^{\bb}) = |\La| g_{\al\bb}(0)\de\sg_1^{\al}\de{\sg'}_1^{\bb}.
\end{equation}
With this inner product, \eqref{Pd_def} becomes:
\begin{equation}\label{Pd_def_1}
    \Pd
    \begin{pmatrix} 
        \de e_{a\g} \\ \de\sg_1^\al 
    \end{pmatrix}
    =
    \begin{pmatrix}
        \bar{g}^{\bb\de}\bar{g}^{\al\g}(-\bar{e}^a_{\bb}\pd_{\al}-\pd_{\al}\bar{e}^a_{\bb}+\pd_{\bb}\bar{e}^a_{\al}) & \frac{\de(\sg)}{\sqrt{\bar{g}(0)}}\de_\al^\de\\
        -\bar{g}^{\al\g}\bar{e}^a_{\al} & 0\\
        -\bar{g}^{\al\g}\ep^a_b \bar{e}^b_{\al} & 0
    \end{pmatrix}
    \begin{pmatrix}
        \de e_{a\g} \\ \de\sg_1^\al
    \end{pmatrix}.
\end{equation}
Multiplying \eqref{P_def_1} by \eqref{Pd_def_1} gives $\Pd P$. The only change with respect to \eqref{PdP_def} is in the upper left entry, where the term  $\frac{\de(\sg)}{\sqrt{\bar{g}(0)}}\de^{\de}_{\bb}$ is added. The same decomposition as in \eqref{dec_PdP}, \eqref{dec_PdP_1} holds, with $c^{\de}_{\bb}$ given by:
\begin{equation}
    c^{\de}_{\bb} = -\frac{1}{2}{\de}^{\de}_{\bb}\pd^2 + \pd^{\de}\Om\pd_{\bb} - \pd_{\bb}\Om\pd^{\de} - {\de}^{\de}_{\bb}\pd_{\al}\Om\pd^{\al} + \frac{\de(\sg)}{\sqrt{\bar{g}(0)}}\de^\de_\bb.
\end{equation} 
Up to the infinite constants $2^{\aleph^2}$, which can be ignored, $J(\bar e,0) = \sqrt{\det(c)}$. 
As in \eqref{c_com_w}, writing $c_\bb^\de$ in complex coordinates gives a simple expression:
\begin{equation}
    [c^\de_\bb]_w
    = e^{-2\Om}
    \begin{pmatrix}
        -\pd\bpd-4\pd\Om\bpd+\de(w) & 0 \\ 0 & -\pd\bpd-4\bpd\Om\pd+\de(w)
    \end{pmatrix},
\end{equation}
and $J(\bar e,0)$ takes the form\footnote{In the function space, multiplication by $e^{-2\Om(\sg)}$ is a multiplication by a diagonal matrix, and so its determinant can be separated.
}:
\begin{equation}
    J(\bar e,0) = |\det\left( e^{-2\Om}\right)| \left|\det(-\pd\bpd-4\pd\Om\bpd+\de(w))\right|.
\end{equation}

% ====================================
\section{Global Gauge Fixing and Boundary Conditions}\label{sec_boun}

In this appendix we discuss the hidden assumptions of \eqref{rel_X_Y_p}, which are related to the boundary conditions at infinity in the case of a non-compact manifold. The asymptotic behavior of the coordinates is defined by \eqref{rel_X_Y_p}: $X^a(\sg) \sim I^a_\al \sg^\al$ as $|\sg|\to\infty$. Hence, the allowed diffeomorphisms $V(\sg^\bb)$ are also limited - any coordinate transformation $\sg'(\sg)$ should satisfy
\begin{equation}\label{p_1}
    \sg'^\al(\sg^\bb) \equiv V^\al(\sg^\bb) = I^{\al}_\bb \sg^\bb + f^\al(\sg^\bb),    
\end{equation}
where $f(\sg)$ has a Fourier transform and is bounded as $|\sg|\to\infty$. With this ``global'' gauge-fixing, coordinate transformations of global rotations and dilatations (as well as all other conformal transformations except translations) are excluded, i.e., in \eqref{p_1} the specific matrix $I^{\al}_\bb$ is chosen, instead of a general constant matrix $A^{\al}_\bb$. In accordance with this exclusion, any vielbein in the $e$ PI will be conformally flat at infinity, i.e., 
\begin{equation}\label{p_2}
    \lim_{|\sg|\to\infty}e^a_\al(\sg) = e^{\bar\Om + \ep\bar\phi} I^a_\al, 
\end{equation}
for some constants $\bar\Om,\bar\phi$. The natural boundary conditions for the vielbein should make the WS flat at infinity, away from the operator insertions. The global rotations and dilatations are thus ``exploited'' to bring the constant vielbein at infinity (which in general, has four degrees of freedom), to the form in \eqref{p_2} (which has two degrees of freedom). In the same spirit, one could also set the vielbein at infinity to $I^a_\al$. However, as was shown in subsection \ref{subsec_reg_IR}, the vielbein are proportional to the identity matrix as a result of the infinite plane limit $|\La|\ATS \to \infty$, so it is not necessary to demand it explicitly.

% ====================================
\section{Fourier Transform of a Power Law}\label{sec_fur}

The large momentum limit led to an integral in \eqref{hml_1}, which is a Fourier transform of a power law:
\begin{equation}\label{fur_1}
    \int d^2\sg \frac{e^{iq\cdot \sg}}{|\sg|^{2\bb}},
\end{equation}
where $\bb \equiv \frac{q^2}{2\pi|\La|} + \Delta$\footnote{In the large momentum limit in subsection \ref{subsec_reg_hml}, where $\Delta << \frac{q^2}{|\La|}$, it was not included when finding the saddle of \eqref{fur_1}. Since it is not necessary to assume this here, 
$\Delta$ will be kept.}. This integral  appears for any CFT when transforming a two-point correlator in position space to momentum space, with $\bb=\Delta$. If $\bb>1$, which happens in particular in the large momentum limit, the integral is divergent because of its the behavior around $\sg=0$. As was explained in subsection \ref{subsec_reg_hml}, one should not include the region $|\sg|<\ep$, avoiding the singularity in \eqref{fur_1}. Then, the integral is solved explicitly, using polar coordinates:
\begin{equation}\label{fur_2}
    \begin{split}
        \int_{\ep<|\sg|} d^2\sg \frac{e^{iq\cdot \sg}}{|\sg|^{2\bb}} &= \int^\infty_{\ep}rdr \int^{2\pi}_0 d\theta \frac{e^{iqr \cos(\theta)}}{r^{2\bb}} = 2\pi \int^\infty_{\ep} r^{1-2\bb} J_0(qr)dr \\  
        &= \pi\Gamma(1-\bb) \left( \frac{1}{\Gamma(\bb)}  \left(\frac{|q|}{2}\right)^{2\bb-2} - \left(\frac{1}{\ep}\right)^{2\bb-2} f_\bb\left(-\frac{1}{4}(\ep|q|)^2\right) \right).
    \end{split}
\end{equation}
Here $J_n(x)$ is the Bessel function of type $J$. The function $f$ is the regularized generalized hypergeometric function:
\begin{equation}
    f_\bb(x) \equiv \frac{_1F_2(1-\bb;\{1,2-\bb\};x)}{\Gamma(1)\Gamma(2-\bb)}.
\end{equation}
The integral is a sum of two contributions. The first does not depend on the regulator, and thus is the desired physical result wanted:
\begin{equation}\label{fur_3}
    \pi\frac{\Gamma(1-\bb)}{\Gamma(\bb)}  \left(\frac{|q|}{2}\right)^{2\bb-2} = \frac{\pi^2}{\sin(\pi\bb)}\frac{1}{\Gamma(\bb)^2}\left(\frac{|q|}{2}\right)^{2\bb-2},
\end{equation}
where the identity $\Gamma(1-x)\Gamma(x)=\frac{\pi}{\sin(\pi x)}$ was used. For a CFT, where $\bb=\Delta$ is independent of the momentum, the momentum dependence is $|q|^{2\Delta-2}$ as expected from naive dimensional analysis of the integral in \eqref{fur_1}. In contrast, in our case, momentum appears also in $\bb$. Then, one has the factor:
\begin{equation}\label{fur_4}
    |q|^{2\bb-2} = |q|^{\frac{q^2}{\pi|\La|}+2\Delta-2},
\end{equation}
which has the exponent of Cardy \eqref{dc_2}. But the momentum also appears in the factors $\frac{1}{\sin(\pi\bb)}$ and $\frac{1}{\Gamma(\bb)^2}$. The $\csc(\pi\bb)$ is odd, since it presents singularities at even positive integer values of $\frac{q^2}{|\La|}$. A similar factor of $\csc(\pi\Delta)$ is present in every CFT, reflecting the fact that for integer dimensions the momentum-space correlator has an extra factor of $\log(q^2\ep^2)$ which diverges when the cutoff is removed. This is an artifact of transforming to momentum space, and this factor disappears in position space. It remains to check the factor $\Gamma(\bb)^{-2}$. In the large momentum limit, $\bb>>1$, the gamma function has the asymptotic form:
\begin{equation}
    \Gamma(x) \xrightarrow[{x \rightarrow \infty}]{} \left( \frac{x}{e} \right)^x \sqrt{\frac{2\pi}{x}},
\end{equation}
which means that:
\begin{equation}\label{fur_5}
    \Gamma(\bb)^{-2} \xrightarrow[{\bb \rightarrow \infty}]{} \left( \frac{e}{\bb} \right)^{2\bb} \frac{\bb}{2\pi} = \left( \frac{2\pi e|\La|}{q^2} \right)^{\frac{q^2}{\pi|\La|}+2\Delta} \frac{q^2}{(2\pi)^2|\La|},
\end{equation}
where to simplify things $\Delta$ was left only in the exponent, i.e., only the first order in $\frac{|\La|\Delta}{q^2}$ was taken and also $\ln\left(\frac{q^2}{\pi|\La|}\right)>>1$ was assumed. One sees that the $\Gamma(\bb)^{-2}$ is going to flip the sign in the exponent of $|q|^{2\bb}$. Plugging \eqref{fur_4},\eqref{fur_5} into \eqref{fur_3} gives finally:
\begin{equation}\label{fur_6}
    \frac{\pi^2}{\sin(\pi\bb)} \frac{\bb}{2\pi} \left(\frac{e}{\bb}\right)^{2\bb} \left(\frac{|q|}{2}\right)^{2\bb-2} = \csc(\pi\bb) \frac{2\pi\bb}{q^2} \left( \frac{e|q|}{2\bb} \right)^{2\bb} \rightarrow 
    \frac{1}{|\La|} \csc(\pi\bb) \left(\frac{\pi e|\La|}{|q|}\right)^{\frac{q^2}{\pi|\La|}+2\Delta}.
\end{equation}
Comparing \eqref{fur_6} to \eqref{hml_4} one finds the same momentum dependence, and also the same dimensional dependence in the form of $|\La|^{-1}$. The only difference is in the numerical coefficient, which is $2/e^{2\Delta}$ compared to $\csc(\pi\bb)$. Such factors cannot be seen using a saddle approximation. Also, as explained before, the $\csc$ is not expected to be physical.

We turn our attention next to the regulator-dependent term in \eqref{fur_2}. It should be removed by an appropriate renormalization. This can be seen by the following. The dimensions are carried outside by the factor $\ep^{2-2\bb}$. The function $f_\bb(x)$ is analytic in $x$. In a CFT, where $\bb$ is independent of the momentum, all the momentum dependence comes from the Taylor expansion of $f$:
\begin{equation}
    f_\bb(x) = \sum_{n=0}^{\infty} f_n(\bb)x^n.
\end{equation}
Using $x=-\frac{1}{4}q^2\ep^2$ gives the series:
\begin{equation}
    \sum_{n=0}^{n=\infty} \left(-\frac{1}{4}\right)^n f_n(\bb) \ep^{2n+2-2\bb}q^{2n}.
\end{equation}
Then, as $\ep|q|\rightarrow 0$ one remains with the divergent terms, from $n=0$ until $n=\floor{\bb}-1$ (recall that the divergence appears for $\bb>1$). The terms with larger values of $n$ vanish as one takes the regulator to zero. The remaining divergent terms are positive integer powers of the momentum, $|q|^{2n}=q^{2n}$. Thus, when transforming back to position space, they give the $2n$-th derivative of $\de(\sg)$. Hence, they are just contact terms which should be subtracted. However, when $\bb$ does depend on the momentum, the dependence of each divergent term on the momentum is complicated - except for the explicit $|q|^n$, the momentum shows up also in $\ep^{-2\bb}$ and in the coefficient $\Gamma(1-\bb)f_n(\bb)$. Moreover, since $\bb$ is very large, one has many divergent terms. The proper  renormalization requires more research. We assume here that the bottom line is still that this term should be removed 

Finally we note that the same result for the Fourier transform of a power law was used in \cite{Asrat:2017tzd} in equation (3.4). The power $4h$ in their notations is $2\bb$ here, so $\bb=2h$. Ignoring the common $D(h)$ in both sides of (3.4), and using their definition for $\g(x)$ in (3.2), one can rewrite their equation as follows:
\begin{equation}
    \begin{split}
        \int d^2x e^{iqx}|x|^{-2\bb} &= \int d^2x e^{iqx}|x|^{-4h} = \pi\g(1-2h)\left(\frac{q^2}{4}\right)^{2h-1} \\ 
        &= \pi \frac{\Gamma(1-2h)}{\Gamma(2h)} \left(\frac{|q|}{2}\right)^{4h-2} = \pi \frac{\Gamma(1-\bb)}{\Gamma(\bb)} \left(\frac{|q|}{2}\right)^{2\bb-2},
    \end{split}
\end{equation}
which is the regulator-independent part in \eqref{fur_2}. They ignored the problems with regularization and renormalization of this Fourier transform, mentioned here, even though their $h$ is also momentum-dependent.

% ====================================
\section{The Large $|\La|$ Limit}\label{sec_hgh_lam}

% ------------------------------------
\subsection{Generalities}

For the large $|\La|$ limit, or the small momentum limit, it is convenient to work in the static gauge, rather than the orthogonal:
\begin{equation}
    \bar X^a(\sg) = I^a_\al\sg^\al.
\end{equation}
In the static gauge the FP determinant is trivial. Considering the constraint 
\begin{equation}
    {\hat C}(V^{\al}(\sg)) = X^a_{\al}(\sg)^{(V^{\al}(\sg))} - I^a_\al\sg^\al,    
\end{equation}
its linearized form around the solution $\bar X$ is:
\begin{equation}\label{P_def_two}
    \begin{split}
        P&\equiv\frac{\de {\hat C}}{\de V},\\
        P \de V^{\al} 
        &= \pd_{\al}\bar X^a
        \de V^{\al} = I^a_\al\de V^\al .
    \end{split}
\end{equation}
Hence, the determinant is an unimportant constant.

In this gauge, the action takes the form:
\begin{equation}
    S_{\TT} = S_0(\psi,g_{\al\bb}) + \frac{\La}{2}\int d^2\sg \ep^{\al\bb}\ep_{ab}(I^a_\al - e^a_{\al})(I^b_\bb - e^b_{\bb}),
\end{equation}
and the PI is:
\begin{equation}
    \frac{1}{Z_{\TT}}\int\DD e\DD\psi\int dA(\sg_1)dA(\sg_2) O(\sg_1)O(\sg_2)e^{iq_1\cdot \sg_1+iq_2\cdot \sg_2-\STT}. 
\end{equation}
Note that the operators we use are not the undeformed operators - rather they are dressed with a factor of the metric:
\begin{equation}
    \OO(X_0) = \sqrt{g(X_0)}O(X_0).
\end{equation}
However, they coincide with the undeformed operators up to order $\La^{-1}$. Alternatively, one can define an alternative operator which indeed coincides with the undeformed operator. The $\sqrt{g}$ factor was present in \eqref{op_def_pos} to make the $d^2\sg$ integral diff-invariant. Instead, one can use the volume form $dX^1\wedge dX^2 = \det(\pd_\al X^a)d^2\sg$, and define:
\begin{equation}\label{op_other_def_pos}
    \OO'(X_0) = \int d^2\sg \det(\pd_\al X^a) O(\sg) \de(X(\sg)-X_0).
\end{equation}
In appendix \ref{sec_diff_op} it will be shown that the large momentum limit of the correlation function of these operators is the same as that of $\OO$.

As $|\La| \to \infty$ one gets a saddle in the $e$ PI at $e^a_\al = I^a_\al$, the WS being identical with the TS, and then one recovers the undeformed partition function and correlator.
\newline
In the next order, $\La^{-1}$, one expands the following quantities:
\begin{equation}
    \begin{split}
        & \de e^a_\al \equiv e^a_\al - I^a_\al \approx \frac{f^a_\al}{\La} + O\left(\La^{-2}\right), \\
        & \OO(X_0) \approx \left(1 + \frac{1}{\La}\Tr(f^a_\al)\right) O(X_0) + O\left(\La^{-2}\right), \\
        & S_0(\psi,g_{\al\bb}) \approx S_0(\psi) + \frac{1}{\La}\int d^2\sg T_0^{\al\bb}(\psi)\de_{ab} I^a_\al \de e^b_\bb + O(\de e^2) =  S_0 + \frac{1}{\La}\int d^2\sg T_0^{\al\bb}\de_{ab} I^a_\al f^b_\bb + O\left(\La^{-2}\right),  \\
        & S_{JT} = \frac{\La}{2}\int d^2\sigma \ep^{\al\bb}\ep_{ab}\de e^a_\al\de e^b_\bb \approx \frac{1}{2\La}\int d^2\sigma \ep^{\al\bb}\ep_{ab} f^a_\al f^b_\bb + O\left(\La^{-2}\right),
    \end{split}
\end{equation}
where the definition for the energy-momentum tensor was used \eqref{def_EMT}. Then, the vielbein PI is a Gaussian integral, where $\DD e = \DD \de e \propto \DD f$:
\begin{equation}
    \begin{split}
        \int \DD f\ e^{-\frac{1}{\La} \left(\int d^2\sg T_0^{\al\bb}\de_{ab} I^a_\al f^b_\bb + \frac{1}{2} \ep^{\al\bb}\ep_{ab} f^a_\al f^b_\bb \right)} = \int \DD f\ e^{-\frac{1}{\La} \int d^2\sg \left(\Tr(fT_0) + \det(f) \right)}
    \end{split}
\end{equation}
The value at the saddle is: $f^a_\al = \ep^{ab}\de_{bc}T_0^{c\nu}\ep_{\nu\al} \rightarrow \Tr(fT_0) + \det(f) = \det(T_0)$.
The partition function is then:
\begin{equation}
    Z_{\TT} \approx \int \DD\psi\  e^{-\left(S_0(\psi) - \frac{1}{\La}\int d^2\sg \det(T_0) \right)}
\end{equation}
and hence the $\TT$-deformation is recovered. From this result, one finds the relation \eqref{INT_5} between the parameters in the $\TT$-formalism and the JT-formalism. Also, since $\Tr(f) = \Tr(T_0)$ vanishes for a CFT, the operators are not modified to first order: $\OO(X_0) \approx O(X_0)$, and the perturbative result is recovered.

% ------------------------------------
\subsection{First Order - $\TT$-Formalism}

We want to compare the leading order correction in our result \eqref{cor_dev_10} to the first order correction in the $\TT$-formalism. (The first order of \eqref{cor_dev_10} will be computed in the next subsection.) The latter is given by expanding the exponent:
\begin{equation}
    e^{\frac{1}{\La}\int d^2\sg \det(T_0)} \approx 1 + \frac{1}{\La}\int d^2\sg \det(T_0),
\end{equation}
giving the following correlator in the undeformed CFT:
\begin{equation}\label{com_fo_1}
    \int d^2\sg \langle O(\sg_1)O(\sg_2)\det(T_0(\sg)) \rangle.
\end{equation}
It is convenient to work with complex coordinates, where the Ward identities can be used in the form:
\begin{equation}
\begin{split}
    & \langle -2\pi T(z) \prod_i O_i(z_i,\bar z_i) \rangle = \sum_j \left(\frac{h_j}{(z-z_j)^2} + \frac{1}{z-z_j}\pd_{z_j} \right)\langle \prod_i O_i(z_i,\bar z_i) \rangle, \\
    & \langle -2\pi \bar T(\bar z) \prod_i O_i(z_i,\bar z_i) \rangle = \sum_j \left(\frac{\bar h_j}{(\bar z - \bar z_j)^2} + \frac{1}{\bar z - \bar z_j}\pd_{\bar z_j} \right)\langle \prod_i O_i(z_i,\bar z_i) \rangle.
\end{split}
\end{equation}
For scalar operators, $h \equiv h_1 = \bar h_1 = h_2 = \bar h_2$.
Writing the determinant with complex variables, and dividing by $4\pi^2$, one gets from \eqref{com_fo_1}:
\begin{equation}
    -4\cdot\frac{1}{4\pi^2} \langle O(z_1,\bar z_1)O(z_2,\bar z_2)T(z)\bar T(\bar z) \rangle = -\frac{1}{\pi^2}\frac{h^2|z_1-z_2|^4}{|z-z_1|^4|z-z_2|^4} \langle O(z_1,\bar z_1)O(z_2,\bar z_2) \rangle.
\end{equation}
Therefore, the first order correction is given by:
\begin{equation}\label{com_fo_2}
    \frac{h^2|\sg_1-\sg_2|^4}{|\La|\pi^2} \langle O(\sg_1)O(\sg_2) \rangle \int d^2\sg\frac{1}{|\sg-\sg_1|^4|\sg-\sg_2|^4},
\end{equation}
where we switched back to Cartesian coordinates. This integral is divergent, and can be evaluated by dimensional regularization to yield the physical term $16\pi\ln(|\sg_{12}|)/|\sg_{12}|^6$ (see (7.14) in \cite{Kraus:2018xrn} and (73) in \cite{He:2020qcs}). 
Let us show this here using point-splitting regularization. Using translational and rotational symmetries of this integral, it can be written as ($\sg^1=x,\sg^2=y,|\sg_{12}|=d$):
\begin{equation}
    \int dxdy\frac{1}{((x+d)^2+y^2)^2(x^2+y^2)^2}.
\end{equation}
The $x$ integral is convergent and gives:
\begin{equation}
    \pi\int_{-\infty}^{\infty} dy \frac{d^2 + 20y^2}{(d^2 + 4y^2)^3}\frac{1}{|y|^3},
\end{equation}
where in the point-splitting regularization $\ep < |y| < \infty$. Expanding in small $\ep$ gives:
\begin{equation}
    \frac{1}{d^6} \left(\pi\frac{d^2}{\ep^2} + 16\pi\ln\left(\frac{d}{\ep}\right) \right),
\end{equation}
and hence the same physical contribution. Inserting it into \eqref{com_fo_2}, and changing $h=\frac{\Delta}{2}$, gives:
\begin{equation}\label{com_fo_3}
    \langle O(\sg_1)O(\sg_2) \rangle^{\La} = \left( 1 + \frac{1}{\pi|\La|}\frac{(2\Delta)^2\ln(|\sg_{12}|)}{|\sg_{12}|^2} \right) \langle O(\sg_1)O(\sg_2) \rangle^0 + O\pr{\La^{-2}}.
\end{equation}

% ------------------------------------
\subsection{First Order - JT-Formalism}

% ------------
\subsubsection{Introduction}

We write again the expressions in \eqref{cor_dev_10}, \eqref{det_dec_3}:
\begin{equation}\label{fo_eval_1}
\begin{split}
    \langle \hat{\OO}(q)\hat{\OO}(-q+k)\rangle & \propto
    \de(k) \int d^2\sg_{21} e^{iq\cdot \sg_{21}} D(z,0) \frac{1}{|\sg_{21}|^{2\Delta}} \left(\frac{\mu|\sg_{21}|}{2}\right)^{-\frac{q^2}{\pi|\La|}}, \\
    D(z,0) & \equiv \frac{\left|\det(- \pd\bpd - 4\pd\Om\bpd + \de(w))\right|}{\left|{\det}'(-(\pd +\pd\ln(\bar z))(\bpd + \bpd\ln(\bar z))\right|}.
\end{split}
\end{equation}
In this expression:
\begin{equation}\label{fo_eval_2}
\begin{split}
    & \bar z = \frac{Qw_{21}}{w(w-w_{21})} - i = -Qf - i, \\
    & e^{2\Om} = |Q|^2 \frac{|w_{21}|^2}{|w|^2|w-w_{21}|^2} - 2i\Re\left( \frac{Qw_{21}}{w(w-w_{21})} \right) - 1 = (Qf + i)(\bar Q \bar f + i), \\
    & f(w;w_{21}) \equiv \frac{1}{w}-\frac{1}{w-w_{21}}
\end{split}
\end{equation}
(where the $\de(\sg_1)$ were used to set $w_1=0$, and $w_2$ is renamed to $w_{21}$)\footnote{Anyway the determinant is expected to be invariant under translations of $w$, so even without the explicit $\de(w_1)$ it could be set to zero.}. When it is clear from the context, the argument of a function will be denoted by the holomorphic variable only. From \eqref{fo_eval_2}, one has for $\pd\ln(\bar z),\pd\Om$ appearing in $D(z,0)$:
\begin{equation}
\begin{split}
    & \pd\Om = -\frac{1}{2}\frac{iQ\pd f}{1 - ifQ}  -\frac{1}{2}\frac{i\bar Q\pd \bar f}{1 - i\bar f \bar Q}, \\
    & \pd\ln(\bar z) = \frac{Q\pd f}{Q\pd f + i}.
\end{split}
\end{equation}
Before we proceed into detailed calculations, it should be noted that the factor $\left(\frac{\mu|\sg|}{2}\right)^{-\frac{q^2}{\pi|\La|}}$ reproduces the same logs as in Cardy's result \eqref{dc_1}. Hence, the evaluation of the determinants in $D(z,0)$ is merely to verify that it does not contribute additional logs\footnote{By the way it is shown to be finite and need not be regulated.}. To first order, the dominant log ($\equiv\ln^1$) is the only meaningful quantity that can be compared, since the free term ($\equiv\ln^0$) depends on the normalization of the operators. After using a fixed normalization, one can compare at second order all three terms ($\ln^2,\ln^1,\ln^0$), but we showed that our operators in the JT-formalism agree with the $\TT$-operators only to first order.

The difficulty in evaluation of \eqref{fo_eval_1} at small momentum is the evaluation of the determinants in $D(z,0)$. The determinants are over all functions' space (one component of a diffeomorphism in the numerator, and a linear combination of $\Om,\phi$ in the denominator). In principle, divergences may appear, and a regularization scale $\ep$ might be needed. Upon expansion in a dimensionful quantity $|Q|$, it may also saturate the dimension of $Q$ (in addition to $w_{21}$). However, divergences did not appear up to second order. Also, it should be noticed that the determinants are invariant under simultaneous rotations of $w_{21}$ and $Q$, and therefore, this invariance will be satisfied for every term in the expansion in powers of $Q$. 

Expansion of the numerator will be considered first. The denominator will have a similar treatment. We define the operators:
\begin{equation}
    A_1 \equiv -\pd\bpd + \de(w),\ 
    B_1 \equiv A_1^{-1}\pr{-4\pd\Om\bpd}.
\end{equation}
$A$ is a real valued operator, $B$ is complex, and hence also $C$. The complex conjugate operators of $B,C$ will be denoted by $\bar B,\bar C$. Using these definitions, the numerator $D_1$ in \eqref{det_dec_3} reads:
\begin{equation}\label{fo_eval_3}
\begin{split}
    & D_1 = |\det(A_1)| \sqrt{ \det(I + B_1)\det(I + \bar B_1) }.
\end{split}
\end{equation}

% ------------
\subsubsection{Fourier Transform Definitions}

For evaluating the traces resulting from the expansion of the determinants in \eqref{fo_eval_3}, a Fourier transform with complex variables will be used. Below, a list of the definitions and formulas needed is given. The momentum $q$ is treated as a contravariant vector. The basic identity is:
\begin{equation}
    \int d^2k e^{iw\cdot k} = N \de(w),
\end{equation}
where $w\cdot q \equiv \de_{\al\bb}w^\al k^\bb = \frac{1}{2}(w\bar k + \bar w k)$ and $N \equiv 4(2\pi)^2$. Using bra-ket notation, the following identities hold in position space:
\begin{align}
    \bra w \ket{w'} = \de(w-w') \ &(Orthonormality) \\
    \hat I = \int dw\ket w\bra w  \ &(Completeness).
\end{align}
In momentum space, a convention needs to be chosen. In general, three relations are involved:
\begin{align}
    \bra k \ket{k'} &= C_1 \de(k-k'), \\
    \hat I &= \int \frac{dk}{C_2}\ket k \bra k, \\
    \bra k \ket{w} &= C_3 e^{\pm i k\cdot w}.
\end{align}
Consistency among them gives: $C_1 = C_2,\ |C_3| = \sqrt{\frac{C_1}{N}}$. The choice for the Fourier transform:
\begin{equation}
    g(w) = \frac{1}{N}\int d^2k e^{iw\cdot k}\Tilde{g}(k) \iff \FF(g)(k) \equiv \Tilde{g}(k) = \int d^2w e^{-iw\cdot k}g(w),
\end{equation}
corresponds to the choice $C_1=N$, and to the minus sign in the exponent in $\bra k\ket{w}$. In particular, with this choice, for the abstract operator $\hat O$ and state $\ket{g}$:
\begin{align}
    \Tilde{g}(k) &\equiv \bra k\ket{g}, \\
    \bra k\hat O\ket{g} &= \int \frac{dk'}{N}O(k,k')g(k'), \\
    \bra k \hat O_1 \hat O_2 \ket{k'} &= \int \frac{dk''}{N}O_1(k,k'')O_2(k'',k'), \\
    \Tr(\hat O) & = \int \frac{dk}{N}O(k,k),
\end{align}
and the relation between the position representation of $\hat O$ and its momentum representation is:
\begin{equation}
    O(k,k') = \int dwdw' \ e^{-ik\cdot w}O(w,w') e^{ik'\cdot w'}.
\end{equation}

% ------------
\subsubsection{First Order Expansion}

Expansion to first order of \eqref{fo_eval_3} involves the following identities:
\begin{equation}
    \sqrt{1 + x} \approx 1 + \frac{x}{2},\qquad     \det(I + B_1) \approx 1 + \Tr(B_1),\qquad
    \pd \Om \approx -\frac{i}{2}(Q\pd f + \bar Q \pd \bar f).
\end{equation}
Therefore:
\begin{equation}
    \sqrt{ \det(I + B_1)\det(I + \bar B_1) } \approx 1 + \frac{1}{2}\pr{\Tr(B_1) + \Tr({\bar B}_1)},
\end{equation}
and:
\begin{equation}
\begin{split}
    \frac{1}{2}\pr{\Tr(B_1) + \Tr({\bar B}_1)} \approx & i\Tr\pr{A^{-1} (Q\pd f \bpd + \bar Q\pd\bar f \bpd + \bar Q\bpd \bar f \pd + Q \bpd f \pd)} = \\ 
    & iQ\Tr\pr{A^{-1}(\pd f \bpd + \bpd f \pd)} + i\bar Q\Tr\pr{A^{-1}(\bpd \bar f \pd + \pd \bar f \bpd)} \equiv \\
    & i(Q\bar w_{21} + \bar Q w_{21})(F_1(|w_{21}|) + F_2(|w_{21}|)),
\end{split}
\end{equation}
where: 
\begin{equation}\label{F_tr}
    F_1(|w_{21}|)\bar w_{21} \equiv \Tr\left(A^{-1}\pd f \bpd\right),\qquad
    F_2(|w_{21}|)w_{21} \equiv\Tr\left(A^{-1}\pd \bar f \bpd\right).
\end{equation}
The other two traces are complex conjugates of these two, as used above. The variables $w_{21},\bar w_{21}$ were pulled out explicitly from the traces, using their rotation symmetry properties. At this stage, it is useful to write the different operators involved in the traces inside $F_1,F_2$ in momentum basis, because $A^{-1}$ has a convenient form in that basis. In the calculations below, the following Fourier transforms were used:
\begin{align}
    & \FF(\pd g(w)) = \frac{i}{2}\bar k\FF(g(w)), \\
    & \FF(g(w-w_2))(k) = \FF(g(w))e^{-iw_2\cdot k}, \\
    & \FF(\de(w)) = 1, \\
    & \FF\left(\frac{1}{w}\right) = -\frac{iN}{2(2\pi)}\frac{1}{k}, \\
    & \FF\left(\frac{1}{w^2}\right) = -\frac{N}{4(2\pi)}\frac{\bar k}{k}, \\
    & \FF\left(\frac{\bar w}{w^2}\right) = \frac{iN}{2(2\pi)}\frac{\bar k}{k^2}.
\end{align}
The abstract operators have the following form in momentum representation:
\begin{equation}
\begin{split}
    \bra{w}\hat O_1\ket{g} \equiv \bpd g(w) & \Rightarrow O_1(w,w') = \bpd \de(w-w') \\ 
    & \Rightarrow  O_1(k,k') = \frac{ik}{2}N\de(k-k'), 
    \\
    \bra{w}\hat O_2\ket{g} \equiv  \pd \bar f(w)g(w) & \Rightarrow O_2(w,w') = 2\pi(\de(w) - \de(w-w_{21})) \\
    & \Rightarrow O_2(k,k') = 2\pi(1 - e^{i(k'-k)\cdot w_{21}}),
    \\
    \bra{w}\hat O'_2\ket{g} \equiv  \pd f(w)g(w) & \Rightarrow O'_2(w,w') = \left(-\frac{1}{w^2} + \frac{1}{(w-w_{21})^2}\right)\de(w-w') 
    \\
    & \Rightarrow O'_2(k,k') = \frac{N}{4(2\pi)}\frac{\bar k' - \bar k}{k' - k}(1 - e^{i(k'-k)\cdot w_{21}}).
\end{split}
\end{equation}
For $A_1$ one has:
\begin{equation}
    \FF(Ag)(k) = \frac{|k|^2}{4}\FF(g)(k) + \frac{1}{4(2\pi)^2}\int d^2k'\FF(g)(k')\footnote{The last term is $g(0)$, so $\FF(Ag)(0) = g(0)$. This addition of $\de(w)$ in position space or $g(0)$ in momentum space makes $A$ invertible. This was the reason for its insertion from the beginning, see appendix \ref{sec_FP}. If $Ag = Af \Rightarrow \FF(Ag) = \FF(Af) \Rightarrow g(0) = f(0)$. Therefore, also $\FF(g) = \FF(f) \Rightarrow g = f$.}.
\end{equation}
Hence:
\begin{equation}\label{fur_A}
    A_1(k,k') = \frac{|k|^2}{4}N\de(k-k') + 1.
\end{equation}
Using this result for the abstract operator $\hat O_3 \equiv \hat A_1^{-1}$ yields:
\begin{equation}
    O_3(k,k') = \frac{4}{|k'|^2}N(\de(k-k') -\de(k)).
\end{equation}
The traces in \eqref{F_tr} can be written as:
\begin{equation}
\begin{split}
    \Tr\left(A_1^{-1}\pd f \bpd\right) &= \int \frac{dk_1dk_2dk_3}{N^3}O_3(k_1,k_2)O'_2(k_2,k_3)O_1(k_3,k_1), \\    
    \Tr\left(A_1^{-1}\pd \bar f \bpd\right) &= \int \frac{dk_1dk_2dk_3}{N^3}O_3(k_1,k_2)O_2(k_2,k_3)O_1(k_3,k_1).
\end{split}
\end{equation}
A short calculation then gives $F_1 = F_2 = -\frac{1}{|w_{21}|^2}$. Therefore, the first order correction to the numerator is:
\begin{equation}
    D_1 \approx |\det\pr{A_1}|\pr{1 -\frac{2iq\cdot \sg_{21}}{\pi|\La||\sg_{21}|^2}}.
\end{equation}

The denominator $D_2$ will have a similar treatment. As in subsection \ref{subsec_PI_vs}, a regulator is introduced to deal with the exclusion of the zero eigenvalue. One has:
\begin{equation}
\begin{split}
    D_2 &= \left|{\det}'(-(\pd +\pd\ln(\bar z))(\bpd + \bpd\ln(\bar z)))\right| = \frac{\left|\det(-(\pd +\pd\ln(\bar z))(\bpd + \bpd\ln(\bar z)) + \ep) \right|}{\ep} \\
    &= \frac{|\det(A_2)|}{\ep} \sqrt{ \det(I + B_2)\det(I + \bar B_2) }
    = |{\det}'(-\pd\bpd)| \lim_{\ep \to 0}\sqrt{ \det(I + B_2)\det(I + \bar B_2) },
\end{split}
\end{equation}
where:
\begin{equation}
    A_2 \equiv -\pd\bpd + \ep, \ 
    B_2 \equiv A_2^{-1}\pr{-\pd\bpd\ln(\bar z) -\bpd\ln(\bar z)\pd -\pd\ln(\bar z)\bpd -\pd\ln(\bar z)\bpd\ln(\bar z)}.
\end{equation}
To first order $\pd\ln(\bar z) \approx -iQ\pd f$ and similarly $\bpd\ln(\bar z) \approx -iQ\bpd f$. Therefore:
\begin{equation}
    \Tr(B_2) \approx iQ\Tr\pr{A_2^{-1} (\pd\bpd f + \bpd f \pd + \pd f \bpd)},
\end{equation}
and the complex conjugate for $\Tr(\bar B_2)$. The abstract operator $\hat O_4$ has the following form in momentum representation:
\begin{equation}
\begin{split}
    \bra{w}\hat O_4\ket{g} \equiv \pd\bpd f(w) g(w) & \Rightarrow O_4(w,w') = 2\pi(\pd\de(w) - \pd\de(w-w_{21})) \\
    & \Rightarrow O_4(k,k') = 2\pi\frac{i(k-k')}{2}(1 - e^{i(k'-k)\cdot w_{21}}),
\end{split}
\end{equation}
For $A_2$ one has:
\begin{equation}
    \FF(Ag) = \frac{|k|^2}{4}\FF(g) + \ep\FF(g),
\end{equation}
hence:
\begin{equation}\label{fur_A2}
    A_2(k,k') = \pr{\frac{|k|^2}{4} + \ep} N\de(k-k').
\end{equation}
Since it is diagonal in momentum representation, its inverse $\hat O_5 \equiv \hat A_2^{-1}$ is trivial:
\begin{equation}
    O_5(k,k') = \frac{4}{|k|^2 + 4\ep} N\de(k-k').
\end{equation}
The traces can be written as:
\begin{equation}
\begin{split}
    \Tr\left(A_2^{-1}\pd\bpd f \right) &= \int \frac{dk_1dk_2}{N^2} O_5(k_1,k_2)O_4(k_2,k_1), \\    
    \Tr\left(A_2^{-1}\pd f \bpd \right) &= \int \frac{dk_1dk_2dk_3}{N^3}O_5(k_1,k_2)O'_2(k_2,k_3)O_1(k_3,k_1), \\
    \Tr\left(A_2^{-1}\bpd f \pd \right) &= \int \frac{dk_1dk_2dk_3}{N^3}O_5(k_1,k_2)\bar O_2(k_2,k_3)\bar O_1(k_3,k_1).
\end{split}
\end{equation}
A short calculation shows that all three vanish. In conclusion, the first order correction to $D(z,0)$ is:
\begin{equation}
    D(z,0) = \frac{D_1}{D_2} \approx \frac{|\det\pr{A_1}|}{|{\det}'(-\pd\bpd)|} \pr{1 -\frac{2iq\cdot \sg_{21}}{\pi|\La||\sg_{21}|^2}}.
\end{equation}
By this our purpose was accomplished to first order - the determinant does not contain logs\footnote{It is also finite and need not be regularized.}.

We want eventually the correlator in position space (in momentum space there may appear unwanted terms due to the Fourier transform, see appendix \ref{sec_fur}). Hence, according to the definition in \eqref{op_def_pos}, we need to Fourier transform back from $q$ to $X$:
\begin{equation}
    \langle \hat{\OO}(X)\hat{\OO}(0) \rangle = \int d^2q e^{-iq\cdot X}\langle \hat{\OO}(q)\hat{\OO}(-q)\rangle.
\end{equation}
This means that every $\sg_{21}$ simply should be replaced by $X$ and each $q$ by $i\nabla_{X}$ (which should act from the left), i.e.:
\begin{equation}
    \langle \hat{\OO}(X)\hat{\OO}(0) \rangle = D(\Om(\nabla_X),0) \frac{1}{|X|^{2\Delta}} \left(\frac{\mu|X|}{2}\right)^{\frac{|\nabla_{X}|^2}{\pi|\La|}}.
\end{equation}
Then, the first order correction for the undeformed correlator is:
\begin{equation}
\begin{split}
    & \frac{1}{\pi|\La|} \left[2\nabla \cdot \left(\frac{X}{|X|^{2\Delta+2}}\right) + \nabla^2 \left(\ln\left(\frac{\mu |X|}{2}\right)\frac{1}{|X|^{2\Delta}}\right) \right] = \\
    & \frac{1}{\pi|\La|}\left(2\pi\de(X) -\frac{8\Delta}{|X|^{2\Delta+2}} + \ln\left(\frac{\mu |X|}{2}\right) \nabla^2 \left(\frac{1}{|X|^{2\Delta}}\right)\right).
\end{split}
\end{equation}
Ignoring the contact term, and calling the zeroth order $C^0(X)$, gives finally:
\begin{equation}
    C^{\La}(X) \approx \left(1 + \frac{1}{\pi|\La|}\left( -\frac{8\Delta}{|X|^2} + \ln\left(\frac{\mu |X|}{2}\right)\nabla^2 \right) \right)C^0(X).
\end{equation}
This has the same dominant log as in \eqref{com_fo_3}. Remembering that in Cardy's notation $2\lambda_c = \frac{1}{\pi|\La|}$, his first order result is also recovered . 

% ====================================
\section{Different Operators}\label{sec_diff_op}

In this appendix we show that the operators defined in \eqref{op_other_def_pos} give the same large momentum limit as \eqref{hml_5}. In momentum space these operators read:
\begin{equation}\label{op_other_def_mom}
    \OO'(q) = \int d^2\sg \det(\pd_\al X^a) O(\sg) e^{iqX(\sg)}.
\end{equation}
Now, the evaluation of the $X$ PI is not as simple as in subsection \ref{subsec_PI_TS}, because of the extra factors of $\det(\pd_\al X^a(\sg_i))$ for each operator. To overcome this difficulty, a source $e^{ib^\al_{ia}\pd_\al X^a(\sg_i)}$ is introduced for each operator. Differentiation with respect to $b^\al_{ia}$ gives the determinant. The new sources will lead to additional divergences, which should be regulated before differentiation. With the sources, the $X$ PI becomes trivial again. The other PI will be the same. It will be beneficial to use complex coordinates, using the following definitions (similar to the charges $Q_i$ \eqref{def_Q}):
\begin{equation}\label{def_B}
\begin{split}
    B^\al_{ia} &= \frac{b^\al_{ia}}{2\pi|\La|}, \\
    B^\al_i &= B^\al_{i1} + iB^\al_{i2} \ ,\ \bar B^\al_i = B^\al_{i1} - iB^\al_{i2}.
\end{split}
\end{equation}
and then:
\begin{equation}
\begin{split}
    B^w &= B^1_1 - B^2_2 + i(B^2_1 + B^1_2), \\
    \bar B^w &= B^1_1 + B^2_2 + i(B^2_1 - B^1_2), \\
    B^{\bar w} &= B^1_1 + B^2_2 + i(B^1_2 - B^2_1) = (\bar B^w)^*, \\
    \bar B^{\bar w} &= B^1_1 - B^2_2 - i(B^2_1 + B^1_2) = (B^w)^*.
\end{split}
\end{equation}
The derivative which leads to $\det(\pd_\al X^a(\sg_i))$ is given by:
\begin{equation}\label{b_eval_4}
\begin{split}
    (-i)^2\left(\frac{\pd^2}{\pd b^1_{i1} \pd b^2_{i2}} - \frac{\pd^2}{\pd b^2_{i1} \pd b^1_{i2}}\right) &= -\frac{1}{4\pi^2|\La|^2}\left(\frac{\pd^2}{\pd B^1_{i1} \pd B^2_{i2}} - \frac{\pd^2}{\pd B^2_{i1} \pd B^1_{i2}}\right) \\
    &= \frac{1}{\pi^2|\La|^2}\left(\frac{\pd^2}{\pd B^w_i \pd \bar B_i^{\bar w}} - \frac{\pd^2}{\pd \bar B^w_i \pd B^{\bar w}_i}\right),
\end{split}
\end{equation}
and it should be evaluated at $B=0$. After evaluating the PI, the expression for the correlator is (the analogue of \eqref{cor_dev_n_1}, without $\sqrt{\bar g(\sg_i)}$, and with modified $\Om,\phi$ and derivatives of $B$ acting on various terms depending on them):
\begin{equation}\label{b_eval_5}
\begin{split}
    & Z_0 |\La|\ATS e^{|\La|\ATS} \de\left(\sn q_i\right) \int \pn d^2\sg_i \de(\sg_1) e^{ i\sum_{i=1}^n q_i\cdot \sg_i}\left\langle \pn O(\sg_i) \right\rangle \int d\bar{\Om}d\bar{\phi} \cdot
    \\
    & \pn\frac{1}{\pi^2|\La|^2}\pr{\frac{\pd^2}{\pd B^w_i \pd \bar B_i^{\bar w}} - \frac{\pd^2}{\pd \bar B^w_i \pd B^{\bar w}_i}}  \left[ D(z,0)\left(\pn e^{-\Delta_i\Om(\sg_i)}\right) e^{ 2\bar\Om + W - |\La|\left(\bar{A} + i\int d^2\sg\bar{e}^a_\al I^\al_a\right)} \right].
\end{split}
\end{equation}
Using:
\begin{equation}
    b^\al_{ia}\pd_\al X^a(\sg_i) = \int d^2\sg b^\al_{ia}\pd_\al X^a(\sg)\de(\sg-\sg_i) = -\int d^2\sg b^\al_{ia} X^a(\sg) \pd_\al\de(\sg-\sg_i), 
\end{equation}
\eqref{OP_PI_1} is modified to:
\begin{equation}\label{b_eval_1}
    \ep^{\al\bb}\ep_{ab}\pd_{\al}\bar{e}^b_{\bb}(\sg) = -\sn\frac{(q_i)_a}{|\La|}\de(\sg-\sg_i) + \sn\frac{b^\al_{ia}}{|\La|}\pd_\al\de(\sg-\sg_i).
\end{equation}
In complex coordinates and 
\eqref{def_B}, \eqref{b_eval_1} reads:
\begin{equation}\label{b_eval_2}
    \begin{pmatrix}
        \pd z \\ \bpd\bar{z}
    \end{pmatrix}
    =
    2\pi\sn
    \begin{pmatrix}
        -\bar Q^i + \bar B_i^w\pd + \bar B_i^{\bar w}\bpd \\ 
        -Q^i + B_i^w\pd + B_i^{\bar w}\bpd
    \end{pmatrix}
    \de(w-w_i),
\end{equation}
with the solution:
\begin{equation}\label{b_eval_3}
    z = e^{\Om + i\phi} = -\sn \frac{\bar Q^i}{\bar w -\bar w_i} + 2\pi\sn \bar B_i^w\de(w-w_i) - \sn\frac{\bar B_i^{\bar w}}{2(\bar w -\bar w_i)^2} + e^{\bar{\Om}+i\bar{\phi}}.
\end{equation}

There are various expressions which should be regularized and calculated before the differentiation with respect to $B$. For the WS area $\bar A$ one integrates $|z|^2$:
\begin{equation}\label{b_eval_6}
\begin{split}
    \bar{A} =& -2\pi\sn \ln(\ep)|Q^i|^2 - 2\pi\cdot2\sum_{i<j} \ln\left({\frac{|\sg_{ij}|}{2}}\right)\de^{ab}Q^i_aQ^j_b + e^{2\bar{\Om}}\ATS \\ 
    & +\frac{\pi}{4\ep^2}\sn |B_i^w|^2 -\frac{\pi}{2}\sum_{i\neq j}\bar B^w_i B^w_j\frac{1}{w^2_{ij}} + (c.c.) \\
    & + \pi\sum_{i\neq j}Q^i\bar B^w_j\frac{1}{w_{ij}} + (c.c.) + \pi\sum_{i\neq j}Q^i\bar B^{\bar w}_j\frac{1}{\bar w_{ij}} + (c.c.) \\
    & + \pi e^{\bar\Om-i\bar\phi}\sn \bar B^w_i + (c.c.).    
\end{split}
\end{equation}
In deriving this expression, point-splitting regularization was used. In particular, the following integrals vanish:
\begin{equation}
\begin{split}
    & \int d^2w \de(w-w_i)\de(w-w_j) = 0 \ (\textrm{every}\ i,j), \\
    & \int d^2w \frac{1}{(\bar w -\bar w_i)^2(w-w_j)^2} = 0 \ (i\neq j), \\
    & \int d^2w \frac{\de(w-w_i)}{w - w_i} = \int d^2w \frac{\de(w-w_i)}{(w - w_i)^2} = 0, \\
    & \int d^2w \frac{1}{(\bar w -\bar w_i)^2} = \int d^2w \frac{1}{(w - w_i)(\bar w -\bar w_i)^2} = 0.
\end{split}
\end{equation}
The other integrals are immediate, except for two, evaluated using polar coordinates:
\begin{equation}
\begin{split}
    & \int \frac{d^2w}{2} \frac{1}{(\bar w -\bar w_i)^2(w-w_i)^2} = \int \frac{d^2w}{2} \frac{1}{|w|^4} = 2\pi \int_\ep^\infty \frac{dr}{r^3} = \frac{\pi}{\ep^2}, \\
    & \int \frac{d^2w}{2} \frac{1}{(w - w_i)(\bar w -\bar w_j)^2} = \frac{1}{\bar w_{ij}} \int \frac{d^2w}{2} \frac{1}{w(\bar w + 1)^2} = \frac{1}{\bar w_{ij}} \int \frac{r dr d\theta}{re^{i\theta}(1 + re^{-i\theta})^2} = \frac{2\pi}{\bar w_{ij}}.
\end{split}
\end{equation}
Another integral appearing in \eqref{b_eval_5} is:
\begin{equation}\label{b_eval_7}
    \int d^2\sg\bar{e}^a_\al I^\al_a = \int \frac{d^2w}{2}(z + \bar z) = 2e^{\bar\Om}\cos(\bar\phi) + \pi\sn\left(\bar B^w_i + B_i^{\bar w}\right).
\end{equation}
The regulation of the divergent term in $e^{-\Delta_i\Om(\sg_i)}$ gives:
\begin{equation}\label{b_eval_8}
    e^{-\Delta_i\Om(\sg_i)} \rightarrow \left(\frac{|Q^i|^2}{\ep^2} + \frac{|B^w_i|^2}{4\ep^4}\right)^{-\frac{\Delta_i}{2}}.
\end{equation}

We need the two derivatives with respect to the $B$'s of:
\begin{equation}\label{b_eval_9}
    D(z,0) e^{-\sum_{i=1}^n \Delta_i\Om(\sg_i) + W -|\La|\left(\bar{A} + i\int d^2\sg\bar{e}^a_\al I^\al_a\right)},
\end{equation}
evaluated at $B=0$. To see what happens, we start with the $B_1$ derivative. The factor $e^{-\Delta_1\Om(\sg_1)}$ contributes when it is differentiated twice or not differentiated at all, so we have:
\begin{equation}\label{b_eval_10}
    e^{-\sum_{i=1}^n\Delta_i\Om(\sg_i)} \Bigg|_{B_1^w = B_1^{\bar w} = 0} \left(\frac{\pd^2}{\pd B^w_1 \pd \bar B_1^{\bar w}} - \frac{\pd^2}{\pd \bar B^w_1 \pd B^{\bar w}_1} -\frac{\Delta_1}{8\ep^2|Q^1|^2} \right)\left[D(z,0) e^{W -|\La|\left(\bar{A} + i\int d^2\sg\bar{e}^a_\al I^\al_a\right)}\right].
\end{equation}
Another term similar to the one proportional to $\ep^{-2}$ will come from the derivatives of $e^{-|\La|\bar A}$. The term $e^{-i|\La|\int d^2\sg\bar{e}^a_\al I^\al_a}$ contributes $-\pi^2|\La|^2$. The derivatives of $W$ and $D(z,0)$ can not be written explicitly, but there is not need to. With the $B$ sources, $W$ changes, but it can evaluated in a similar manner to the calculation in subsection \ref{subsec_reg_UV}, using point-splitting regularization. When $B \neq 0$, the term $B_i^w/(w-w_i)^2$ in $\bar z$ is more singular than $Q^i/(w-w_i)$, and it will dominate the evaluation of $W$. For:
\begin{equation}
    W = \frac{c}{48\pi} \int d^2w \pr{\frac{\bpd\bar z \pd \bar z}{\bar z^2} + \frac{\pd z \bpd z}{z^2} + \frac{\pd z \bpd\bar z}{|z|^2} + \frac{\bpd z \pd\bar z}{|z|^2}},
\end{equation}
the first two integrals give $-16\pi n$. The third integral vanishes in our regularization. The fourth integral splits into a finite part $I_0$ and to a regulator-dependent part $I_1$, as in \eqref{Weyl_4}. The difference from the $B=0$ case, is that now $z = 0$ has $2n$ roots, hence $2n$ (simple) poles for $\frac{\pd z}{z}$. The numerator still has only $n$-poles, at $\{w_i\}$.\footnote{The contour integral at $|w| \to \infty$ vanishes, as $\frac{\pd z}{z} \to w^{-3}$. Therefore the sum of its residues should be zero. Indeed, the residue at each pole coming from the denominator equals $1$, and from the numerator it is $-2$.} Altogether, one finds:
\begin{equation}
    I_1 = 4\sum_{i=1}^n 4\pi \ln\pr{\frac{r}{\ep}} + \sum_{i=1}^{2n} 4\pi \ln\pr{\frac{r}{\ep}} = 24\pi\ln\pr{\frac{r}{\ep}},
\end{equation}
and it is independent on $B$. Therefore, \eqref{b_eval_10} turns into:
\begin{equation}
\begin{split}
    & D(z,0) e^{-\sum_{i=1}^n\Delta_i\Om(\sg_i) + W -|\La|\left(\bar{A} + i\int d^2\sg\bar{e}^a_\al I^\al_a\right)} \Bigg|_{B_1^w = B_1^{\bar w} = 0} \cdot \\
    & \pr{ -\frac{\Delta_1}{8\ep^2|Q^1|^2} -\frac{\pi|\La|}{4\ep^2} -\pi^2|\La|^2 + \pr{\frac{\pd^2}{\pd B^w_1 \pd \bar B_1^{\bar w}} - \frac{\pd^2}{\pd \bar B^w_1 \pd B^{\bar w}_1}} \pr{ \ln(D(z,0)) + I_0}}.
\end{split}
\end{equation}
The determinant $I_0$ does not depend on $\ep$, and is negligible compared to the first two terms in the brackets, and the same is true for $\ln(D(z,0))$. 

Acting with all $B$-derivatives on \eqref{b_eval_9} gives:
\begin{equation}
    \left[ \pn \pr{-\frac{1}{\ep^2}} \left(\frac{|Q^i|}{\ep}\right)^{-\Delta_i} \left( \frac{\pi|\La|}{4} + \frac{\Delta_i}{8|Q^i|^2} \right) \right] \left[D(z,0) e^{-\sum_{i=1}^n\Delta_i\Om(\sg_i) + W -|\La|\left(\bar{A} + i\int d^2\sg\bar{e}^a_\al I^\al_a\right)}\right],
\end{equation}
where all quantities are evaluated at $B=0$, and thus equal to their value before. Hence the difference from $\OO$ is the new factor, $-\frac{1}{\ep^2}\pr{\frac{\pi|\La|}{4} + \frac{\Delta_i}{8|Q^i|^2}}$, and the power of $\frac{|Q^i|}{\ep}$ changed from $2-\Delta_i$ to $-\Delta_i$. As these two factors are polynomials in the momenta, it is clear that the large momentum limit, which is dominated by the exponential dependence in the momentum in $e^{i\sum_{i=1}^n q_i\cdot \sg_i - |\La|\bar A}$, will not change.
It is also clear what is the appropriate renormalization for these operators. First, the same factor of $(\mu\ep)^{-2\pi|\La||Q^i|^2}$ is required to eliminate the divergence in $\bar A$. Next, $\left(\frac{\ep}{|Q^i|}\right)^{-\Delta_i}$ is needed, as well as multiplication by $\ep^2$ for each operator. The question is which quantity will make $\ep^2$ dimensionless. If the correlator in position space is expected to decay at large distances, negative powers of momentum can not appear. Hence, $-\ep^2|Q^i|^2|\La|^2$ is appropriate. A factor $|\La|^2$ is needed to cancel the $\frac{1}{|\La|^2}$ in front of each $B_i$ derivative in \eqref{b_eval_4}, making the large $|\La|$ limit independent of $|\La|$. Putting everything together:
\begin{equation}
    \OO'_i(q_i) \rightarrow \hat{\OO}'_i(q_i) \equiv (\mu\ep)^{-2\pi|\La||Q^i|^2} \left(\frac{\ep}{|Q^i|}\right)^{-\Delta_i} \pr{-\ep^2|Q^i|^2|\La|^2} \OO'_i(q_i).
\end{equation}
The correlator of the renormalized operators will be (after $|\La|\ATS \to \infty$):
\begin{equation}
\begin{split}
    Z_0\pi \de\left(\sn q_i\right)
    \int & \pn d^2\sg_i \de(\sg_1) e^{i\sum_{i=1}^n q_i\cdot \sg_i} C^0(\{\sg_i\}) D(z,0) e^{\frac{cI_0}{48\pi}} \cdot
    \\
    & \pn \left(\frac{|\La||Q^i|^2}{4\pi} + \frac{\Delta_i}{8\pi^2} \right)
    \prod_{i<j} \left( {\frac{\mu|\sg_{ij}|}{2}} \right) ^{4\pi|\La|\de^{ab}Q^i_aQ^j_b}.
\end{split}
\end{equation}
Comparing to \eqref{UV_6}, the new factor of $\left(\frac{|\La||Q^i|^2}{4\pi} + \frac{\Delta_i}{8\pi^2} \right)$ will make the two correlators different in the small momentum limit, as indicated in appendix \ref{sec_hgh_lam}. It will not change the coefficient of the logarithm at order $|\La|^{-1}$ if one further modifies the renormalized operators by the factor $\frac{8\pi^2}{\Delta_i}$.

\end{document}